\documentclass[fleqn,usenatbib]{mnras}

\usepackage{newtxtext,newtxmath}

\usepackage[T1]{fontenc}

\DeclareRobustCommand{\VAN}[3]{#2}
\let\VANthebibliography\thebibliography
\def\thebibliography{\DeclareRobustCommand{\VAN}[3]{##3}\VANthebibliography}
\usepackage{natbib}

\usepackage{graphicx}	
\usepackage{amsmath}	
\usepackage{lscape} 
\usepackage{threeparttable} 

\def\aj{AJ}
\def\apj{ApJ}
\def\apjl{ApJL}
\def\apjs{ApJS}
\def\ao{Appl.~Opt.}
\def\aap{A\&A}
\def\aapr{A\&A~Rev.}
\def\mnras{MNRAS}
\def\pasa{PASA}
\def\pasp{PASP}
\def\ssr{Space~Sci.~Rev.}
\def\nat{Nature}
 
\DeclareMathAlphabet{\mathsc}{OT1}{cmr}{m}{sc} 
\def\testbx{bx}%
\DeclareRobustCommand{\ion}[2]{%
\relax\ifmmode 
\ifx\testbx\f@series 
{\mathbf{#1\,\mathsc{#2}}}\else 
{\mathrm{#1\,\mathsc{#2}}}\fi 
\else\textup{#1\,{\mdseries\textsc{#2}}}%
\fi} 

\newcommand{\Nai} {\ion{Na}{i}}

\newcommand{\Oia} {[\ion{O}{i}]}

\newcommand{\ebv}{\mbox{$E(B-V)$}} 
 
\newcommand{\degree}{\mbox{$^\circ$}}

\newcommand{\kms}{\mbox{$\rm{\,km\,s^{-1}}$}}

\newcommand{\mum}{\mbox{$\mu{\rm m}$}}

\title[The nature of AT2020ohl]{AT2020ohl: Its nature and probable implications}

\author[Roy et al.]{
Rupak Roy,$^{1}$\thanks{E-mail: rupakroy1980@gmail.com, rupak.roy@manipal.edu}
Samir Mandal,$^{2}$
D. K. Sahu,$^{3}$
G. C. Anupama,$^{3}$
Sumana Nandi$^{1}$
and Brijesh Kumar$^{4}$
\\\\
$^{1}$Manipal Centre for Natural Sciences, Manipal Academy of Higher Education, Manipal - 576104, Karnataka, India\\
$^{2}$Indian Institute of Space Science \& Technology, Trivandrum - 695547, Kerala, India\\
$^{3}$Indian Institute of Astrophysics, II Block Koramangala, Bangalore -  560034, India\\
$^{4}$Aryabhatta Research Institute of Observational Sciences, Manora Peak, Nainital - 263001, Uttarakhand, India\\
}

\date{Accepted 2024 January 26. Received 2024 January 02; in original form 2023 July 12}

\pubyear{2015}

\begin{document}
\label{firstpage}
\pagerange{\pageref{firstpage}--\pageref{lastpage}}
\maketitle

\begin{abstract}
 ASASSN-20hx, a.k.a AT2020ohl, is an ambiguous nuclear transient (ANT), which was discovered in the nearby galaxy NGC6297 by the All-Sky Automated Survey for
 Supernovae (ASAS-SN). We have investigated the evolution of AT2020ohl using a
 multi-wavelength dataset
 to explain the geometry of the system and the energy radiated by it between
 X-ray and radio wavelengths. Our X-ray, UV/optical, and radio observations of the 
 object jointly clarify the association of AT2020ohl with the nuclear activity of 
 NGC6297. We detected radio counterpart of AT2020ohl 111 
 days and 313 days after the discovery in JVLA X-band with flux densities 47$\pm$14 
 $\mu$Jy and 34$\pm$3 $\mu$Jy, respectively. Using multi-wavelength data analysis,
 we nullify the possibility of associating any
 stellar disruption process with this event. We found some evidence showing that the host galaxy 
 is a merger remnant, so the possibility of a binary SMBH 
 system can not be ruled out. The central SMBH has a mass
 of $\sim1.2\times10^7$ M$_\odot$. We propose the accretion 
 disk activity as the origin of AT2020ohl $-$ it is either 
 due to disk accretion event 
 onto the central SMBH or due to the sudden accretion activity in a preexisting accretion disk of the system 
 during the interaction of two SMBHs which became gravitationally bound during a merger 
 process. However, we also admit that with the existing
 dataset, it is impossible to say definitively, among these two probabilities, which one is the origin of this nuclear transient.
\end{abstract}




\section{Introduction}\label{sec:intro}
 The region ($\lesssim100$pc) close to the center of any 
 galaxy is expected to be the harbor of various cosmic 
 catastrophes, although it was hard to detect them 
 observationally even a few decades ago. However, with the advent of all-sky
 survey programs in the recent past (e.g., Panoramic Survey Telescope \& Rapid
 Response System (Pan-STARRS), \citealt{2004AN....325..636H};
 Texas Supernova Search, \citealt{2006PhDT........13Q}; Palomar Transient
 Factory (PTF), \citealt{2009PASP..121.1334R}; Catalina Real-time Transient
 Survey (CRTS), \citealt{2009ApJ...696..870D}; All-Sky
 Automated Survey for SuperNovae (ASAS-SN), \citealt{2014AAS...22323603S};
 Zwicky Transient Facility (ZTF), \citealt{2019PASP..131a8002B})  
 highly energetic explosions/flares have been discovered at the centers of many
 nearby and distant galaxies.
 Multi-channel origin of these explosions is quite obvious. Among
 these energetic phenomena, if we consider only those with peak radiated 
 power more than 10$^{41}$ erg\,s$^{-1}$, there are three 
 known physical processes responsible for these events. One possibility is
 the occurrence of luminous supernova (SN) from very massive stars located
 within $\sim100$ pc from the center (cf., \citealt{2017NatAs...1..865K}). The
 supernova remnant Sgr-A East located within
 $\sim2$ pc from the center of our Milky Way galaxy (Sgr-A$^*$) is the nearest
 evidence in this regard (cf., \citealt{2002ApJ...570..671M} and references
 therein). The second possibility is the disruption of a star (roughly a solar
 mass) during its sufficiently close approach to the Supermassive Black Hole
 (SMBH) at the center of a galaxy. This process, which was theoretically
 predicted more than 40 yrs ago (cf., \citealt{1975Natur.254..295H,
 1988Natur.333..523R}) although observationally found much later (cf.,
 \citealt{1999A&A...343..775K, 2009ApJ...698.1367G}), is known as a Tidal
 Disruption Event (TDE). Both possibilities are connected to
 disruptions of stars, although the underlying physics is 
 completely different. The third channel is related to the 
 accretion phenomena onto the central SMBHs of the Active Galactic Nuclei (AGN) and the 
 instability associated with these processes (cf., \citealt{2012agn..book.....B,
 2017A&ARv..25....2P}).

 However, observationally distinguishing the 
 electromagnetic features of these 
 nuclear transients is 
 still
 a challenge. 
 As a consequence, the
 nature of several newly discovered nuclear transients are debated.
For example, (1) the same event has been
 interpreted early as an SN and later as a TDE or vice versa (e.g., ASASSN-15lh,
 \citealt{2016Sci...351..257D, 2016NatAs...1E...2L}; CSS100217,
 \citealt{2011ApJ...735..106D, 2017ApJ...843..106B}; ASASN-17jz,
 \citealt{2022ApJ...933..196H} and references therein), (2) a few nuclear events
 have simultaneously shown TDE and AGN-like features, and
 in some cases detailed study could not fix their origin. 
 These are now called as Ambiguous Nuclear Transients (ANT) 
 (e.g., 1ES 1927+654/ASASSN-18el,
 \citealt{2019ApJ...883...94T}; ASASSN-18jd, \citealt{2020MNRAS.494.2538N}),
 (3) a few of them did not show any significant spectral changes during their
 evolution (e.g., Dougie \citealt{2015ApJ...798...12V}; ASASSN-20hx,
 \citealt{2022ApJ...930...12H}), making their nature questionable.
 Some of these ambiguous phenomena may be 
 related to the intrinsic
 properties of the central SMBH itself, especially associated with its
 dynamics and recurrent activity, which are not yet well understood.
 Discoveries have confirmed the presence of binary SMBH with inner separations as
 small as few pc (\citealt{2017NatAs...1..727K} and references therein) at the
 center of a merged system. Merging (or completely merged) galaxies and/or gas 
 accumulation processes in a single, binary (or trinary) system of galaxies is a
 phenomenon with a time scale of millions of years. These are believed to
 be switching processes between inactive and active states of the SMBHs (e.g.,
 \citealt{2001A&A...374..861S, 2019MNRAS.486.5158N} and references therein).
 However, for a given system, how these processes are initiated is
 yet unknown. Needless to say, advanced transient survey programs have provided
 a unique opportunity to probe the variety of SMBH triggering mechanisms and to 
 search for their different accretion behaviours. Apart from the regular flaring
 activities of AGNs (which are more prominent in 
 its Blazar subclass), long-term
 spectral and temporal evolutions have also been noticed in active SMBH
 systems (e.g., `Changing Look' quasar, \citealt{2015ApJ...800..144L}). Recent
 observations of transients also demand the existence of `rejuvenated' SMBHs
 (e.g., AT2017bgt, \citealt{2019NatAs...3..242T}). While 
 `Changing 
 Look' behaviours may be due to a significant 
 change in the accretion rate onto the SMBH over 
 a longer timescale, or 
 due to the dust attenuation along the line of the
 sight; the `rejuvenated' scenario has been claimed to be due to a sudden matter
 flow in the disk of active SMBH, causing a several-order-of-magnitude rise in
 the ultraviolet (UV) and X-ray flux, along with the appearance of Bowen fluorescence lines. 
 Therefore, the `rejuvenated' SMBH scenario 
 is expected to be observable in AGN dominated 
 systems.
 
 However, the occurrence of transients due to 
 `rejuvenated' SMBH, in inactive or extremely 
 mildly active systems,
 or association of such transient activity with the 
 dynamics of a binary SMBH in a merged system is as yet 
 unknown. In this work, revisit the nuclear transient 
 event AT2020ohl/ASASSN-20hx, which was hosted by an 
 mildly-active galaxy that might be a merger remnant.
 The event was extensively followed by \citet{2022ApJ...930...12H}. The evolution of the transient in X-ray, near-ultraviolet (NUV) and optical wavelengths was reported, starting from $-$30 to 275 days relative to the maximum UV/optical brightness. The observed UV/optical lightcurves were fitted with TDE-models, although the shallow post-maximum decline of the lightcurves and non-thermal X-ray spectrum 
 were indicators of the non-TDE origin of the object. However, it could not be characterized as a Tidal Disruption Event (TDE),
 or phenomena associated with Active Galactic Nuclei (AGN).
 Here we present X-ray, near-ultraviolet (NUV), optical imaging and spectroscopic, and radio
 photometric 
 observations of AT2020ohl and the properties
 of its host galaxy. The NUV and optical photometric 
 data cover about 800 days
 since the discovery. The pre-transient optical monitoring of the object of
 a similar time span is also been presented in this work.   

 Section \ref{sec:obs} describes the discovery and 
 follow-up of the transient at
different electromagnetic (EM) wavelengths. The nature of
 the host galaxy, the extinction along the line of sight 
 and the distance to the system are discussed in section 
 \ref{sec:host}. In section \ref{sec:evol}, the 
 evolution of the transient in
 different EM wavelengths is described. The 
 interpretations of our
 observational results are discussed in the section 
 \ref{sec:discuss} and conclusions are drawn in the same section.  

\section{AT2020ohl $-$ Discovery \& follow-up}\label{sec:obs}
 AT2020ohl/ASASSN-20hx was discovered by 
 the ASAS-SN survey \citep{2020ATel13891....1B} on UT 2020-07-10.34 at g$\sim$16.7 mag at
 the center of the nearby (z=0.0167) galaxy NGC 6297. 
 It was initially classified as a TDE \citep{2020ATel13893....1H}, and further analysis of the
 early Neil Gehrels {\it Swift}\footnote{{\it Neil Gehrels Swift Gamma-ray Burst
 Mission} (hereafter {\it Swift}), \citealt{2004ApJ...611.1005G}}/XRT
 observations
 suggested it as a hard X-ray TDE candidate with a spectral index roughly between 2.2$-$2.6
 \citep{2020ATel13895....1L}.
 The transient was extensively followed by \citet{2022ApJ...930...12H}, mainly in the 
 X-ray, near-UV (NUV), optical bands, and mid-IR in a few epochs. They
 also analysed the early data observed by TESS\footnote{Transiting Exoplanet
 Survey Satellite, \citealt{2015JATIS...1a4003R}} during the rising phase of the
 transient. The rise in the TESS data was estimated to be on JD =
 2459023.3$^{+0.8}_{-0.6}$ \citep{2022ApJ...930...12H}, compared to the 
 discovery date i.e.,
JD = 2459040.8, 
 17.5 days later 
 the initial explosion. This
 value is also consistent with the limit of non-detection by
 \citet{2020ATel13891....1B}. In the present work, we use JD = 2459023.3 
as the epoch of explosion (t$_0$), and all the phases are measured with respect to
 t$_0$.

 Although the early bluish spectral energy distribution (SED) of the source, 
 starting from UV to optical, along with X-ray detection, supported its
 TDE nature 
 (as discussed in the subsequent sections, and
 also by \citealt{2022ApJ...930...12H}), the overall behaviour of AT2020ohl was
 not like canonical TDEs. Therefore detailed multi-wavelength follow-ups and
 analyses were necessary. To achieve this goal, we performed targeted
 observations of the transient in the X-ray, UV, Optical, and Radio bands. A brief
 overview of our observations has been discussed below. 

\subsection{Optical Spectroscopic Observation}\label{sec:obs:spec}
 Long-slit low-resolution optical spectra  ($0.4-0.8 \mu$m) were
 collected from the 2m Himalayan Chandra Telescope (HCT) at 7 epochs, viz.
 2020-08-31, 2020-09-08, 2020-09-29, 2020-10-16, 2020-12-02, 2021-03-27, and
 2021-10-19. A journal of spectroscopic observations is given in Table
 \ref{tab:speclog}. The data were reduced in the IRAF\footnote{IRAF stands
 for Image Reduction and Analysis Facility distributed by the National Optical Astronomy
 Observatory, which is operated by the Association of Universities for Research
 in Astronomy, Inc., under a cooperative agreement with the National Science
 Foundation.} environment. Bias and flat-fielding were performed on all the
 frames. Cosmic-ray rejection was done using the Laplacian
 kernel detection method \citep{2001PASP..113.1420V}. The instrumental 
 resolution (FWHM) of these spectra, as measured from the \Oia\,$\lambda$5577\ emission
 skyline, was found to lie between 6 and 10\AA\ ($\sim 322-510$ \kms). Flux
 calibration was done using standard spectrophotometric star fluxes from
 \citet{1994PASP..106..566H}. SDSS survey\footnote{\url{https://skyserver.sdss.org/dr12/en/tools/chart/navi.aspx}} provides a low-resolution pre-explosion spectrum with moderate SNR of 
 the center of NGC 6297.
 These archival data of the host are used in our analysis.

\subsection{JVLA Radio Observation}\label{sec:obs:radio}
 Prior to AT2020ohl, the center of NGC6297 was not detected in radio wavebands.
 We analysed available archival radio data from the 
 FIRST\footnote{FIRST stands
 for Faint Images of the Radio Sky at Twenty-cm \citep{1995ApJ...450..559B}.},
 NVSS\footnote{NVSS stands for The NRAO VLA Sky Survey
 \citep{1998AJ....115.1693C}.}, and GB6\footnote{GB6 is the Green Bank telescope
 4.85 GHz sky survey \citep{1996ApJS..103..427G}.} radio sky surveys. The host
 was detected in neither the 1.4 GHz NVSS or  FIRST surveys (with limiting sky-rms
 0.45 and 0.15 mJy/beam, respectively), nor in the 5 GHz GB6 survey (limiting sky-rms
 3 mJy/beam). We did not find any pre-transient radio emission in the 3 GHz radio
 sky survey VLASS\footnote{VLASS stands for VLA Sky Survey
 \citep{2020PASP..132c5001L}.} as well (limiting sky-rms 0.12 mJy/beam).

 Motivated by the detection of the X-ray counterpart of AT2020ohl, we observed the
 field using Jansky Very Large Array (JVLA) at three epochs in two frequencies 
 $-$ X-band (8$-$12 GHz), and C-band (4$-$8 GHz). The X-band
 observations were done on 2020-10-29 (proposal code VLA/20B-427) and 2021-05-19 (proposal code VLA/21A-420) using the  B-array and D-array configurations, 
 respectively. In C-band, it was done only in the D-array configuration on
 2021-05-24 ( VLA/21A-420). The raw data were calibrated using the standard VLA
 pipeline
 tool\footnote{\url{https://science.nrao.edu/facilities/vla/data-processing/pipeline}}
 which is based on CASA\footnote{\url{https://casa.nrao.edu/}}. 
 After
 calibration the target data was extracted using the task \texttt{split}, and then imaged by \texttt{tclean}. 
 The JVLA Synthesized Beam width 
 in the B-array in X-band is $0\farcs6$, whereas in D-array, they are $7\farcs2$ and 12\arcsec for the X-band and C-band, respectively. A radio counterpart was detected from NGC
 6297 in all of our observations. Details of the radio observations are given in
 Table \ref{tab:vladata}.

 Due to its higher spatial resolution, the X-band observation on 2020-10-29 revealed
 two radio sources near the center of NGC6297 $-$ a central component with a second radio 
 knot at a separation of $1\farcs5$ from the center of the galaxy. This angular
 separation is extremely small compared to the projected dimension (of major
 axis\footnote{Noteworthy, the length of the major axis of a galaxy, projected
 on the sky-plane is its length measured at the isophotal level 25
 mag/arcsec$^2$ in the B-band.}) of the host
 \citep{2014A&A...570A..13M}\footnote{\label{leda}\url{http://leda.univ-lyon1.fr}},
 which is roughly 38\arcsec. However the aforementioned high resolution
 radio observation was inadequate to distinguish the transient from those two
 radio counterparts at the center of NGC 6297.
 Noteworthy, the optical spectra (acquired using a slit of width $1\farcs5$) also contain both host and transient spectra. Therefore the present data set cannot spatially resolve the transient from its host's centre. Nevertheless, since the X-ray
 counterpart of AT2020ohl was detected by {\it Swift}/XRT, high-resolution X-ray
 imaging of the field using CHANDRA X-ray
 observatory\footnote{\url{https://chandra.harvard.edu/}} was done to mark the
 transient position precisely. 

 \subsection{{\it Swift}/UVOT Observation}\label{sec:obs:uvopt}
 {\it Swift}/UVOT data have also been used to study the near-UV (NUV) and
 optical counterparts of the transient\footnote{Jason T. Hinkle is the Principal Investigator of all of the {\it Swift}/UVOT and XRT observations. The raw public data of the entire observation is available in the Heasarc website (\url{https://heasarc.nasa.gov/}).}. 
 Early {\it Swift}/UVOT observations (till 275 days post-maximum) were reported by 
 \citet{2022ApJ...930...12H}. In this work, we have compiled the {\it Swift} NUV data till $\sim$800 days post-maximum. The public {\it Swift} data is available 
 in the website of `Heasarc'.
 The {\it Swift} data were reduced by using the standard
 pipeline available in the \texttt{HEAsoft} software
 package\footnote{\url{https://heasarc.nasa.gov/lheasoft/}}. 
 UVOT observations at each epoch were conducted using one or several
 orbits. To improve the signal-to-noise ratio (SNR) in a given band at a
 particular epoch, all orbit data of that corresponding epoch have been co-added
 using the \texttt{HEAsoft} routine \texttt{uvotimsum}. The routine
 \texttt{uvotdetect} is used to determine the correct position of the
 transient (which is consistent with the ground-based optical observations). We
 used the routine \texttt{uvotsource} to 
 perform aperture photometry. For source extraction, a small
 aperture of radius $3\farcs5$ was used to minimize the host
 contamination, while an aperture of radius 50$\arcsec$ was used to determine
 the background\footnote{`Aperture-correction' has been applied to the count rates to correct for the missing counts while computing the photometry. 
 It is possible that
 when multiple photons are coincident on the same area of the UVOT detector, it
 does not count all the photons. This effect is known as `coincidence loss'. The
 \texttt{uvotsource} routine makes the necessary correction to compensate for
 the `coincidence loss' effect. Noteworthy, to remove the
 host contamination from the measured NUV-optical magnitudes,
 pre-flare photometric and spectroscopic data of the galaxy taken from $GALEX$
 and SDSS has been used. UVOT also imaged the host in {\it uvw1} and
 {\it U} bands on 2018-07-27 and 2019-06-21, much before the transient happened,
 as part of the
 Swift Gravitational Wave Galaxy Survey (SGWGS, \citealt{2012ApJS..203...28E}).
 These UVOT images were also used to compute the host magnitudes in those bands,
 using an identical source region. The method adopted to
 measure the central magnitudes of the host in different UVOT bands is discussed 
 in \S\ref{sec:host}.}. The host-subtracted UVOT photometry is
 presented in Table \ref{tab:phot_uvot}.

\begin{figure*}
\centering
\includegraphics[width=8.5cm,angle=0]{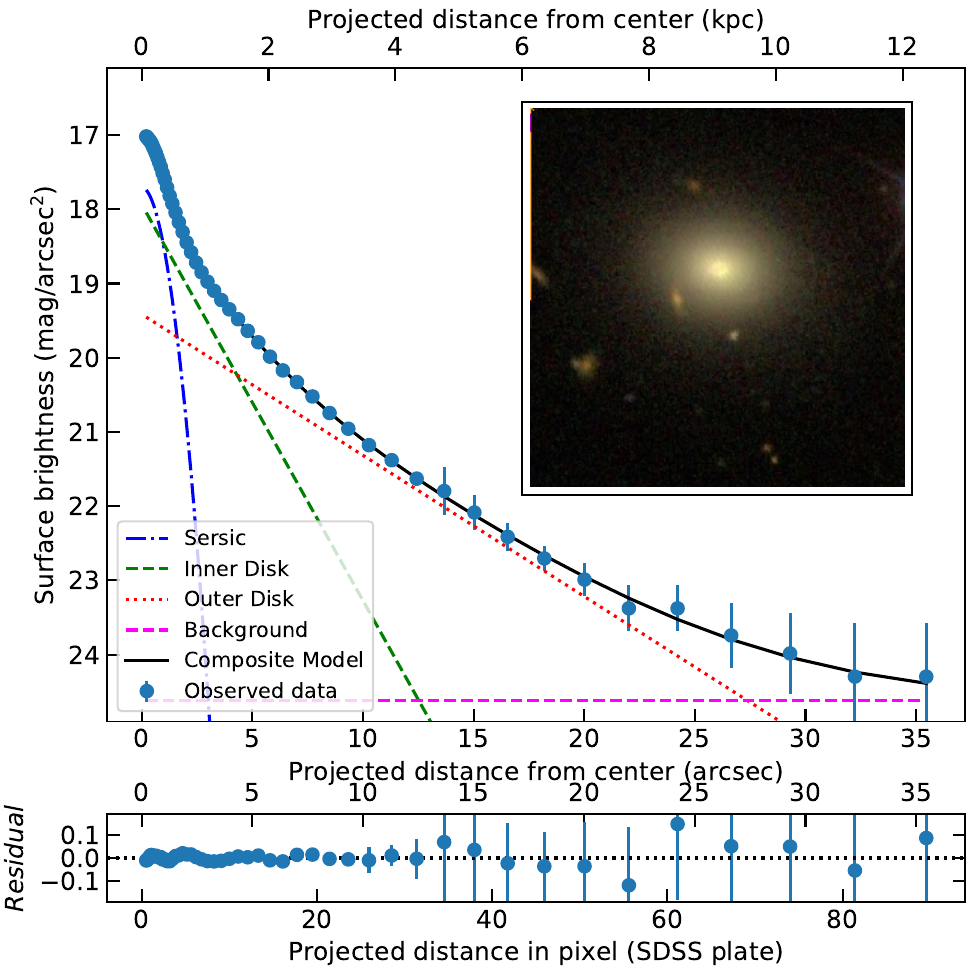}
\hskip 0.5cm
\includegraphics[width=8.5cm,angle=0]{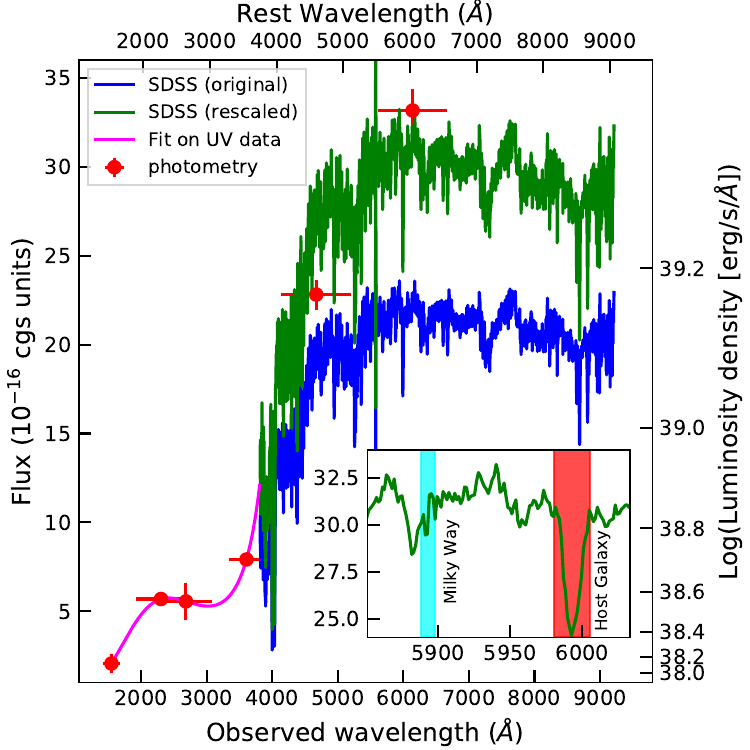}
\caption{NGC6297 $-$ the host of AT2020ohl. {\bf Left Panel:} Surface brightness
 of the galaxy as a function of radius
 (blue points). The black curve shows the resulting model fit for the SDSS $r$-band
 data
 from the combination of Sersic (blue dash-dot line), inner
 exponential-disk (green dash line), and outer exponential-disk (red dotted
 line) profiles. The 
 The dotted-dashed purple line is the background, while the inset shows a $\sim1.5\arcmin\times1.5\arcmin$ SDSS colour composite image of NGC6297.
 {\bf Right Panel:} The UV-optical spectrum of the center of NGC6297. The 
 blue spectrum shows the original {\bf flux calibrated} SDSS observation 
 (using an optical 
 fiber of radius $1\farcs5$). The green spectrum is obtained after 
 re-scaling the SDSS spectrum with respect to the central `g' and `r' 
 band fluxes of the host computed within an aperture of radius $3\farcs5$ 
(red points). These photometric measurements
 have been done on the pre-flare SDSS images. The magenta curve is the low-order
 polynomial fit over the available pre-flare UV data. The inset shows the region
 around the \Nai\,D$\lambda\lambda5890,5896$ doublet in the calibrated observed
 spectrum. The cyan shaded region shows the contribution due to the Milky Way, while
 the red region marks that due to the host galaxy.}
\label{fig:host}
\end{figure*}
 
\subsection{{\it Swift}/XRT Observation}\label{sec:obs:xrt}
 {\it Swift} X-ray Telescope (XRT) operates in the energy range 0.3 - 10 keV 
 \citep{2005SSRv..120..165B}. XRT monitored the source AT2020ohl over 800 days, 
 starting from
 2020-07-19 (MJD 59049). Like NUV observations, we have also used the publicly available {\it Swift}/XRT data of this event till $\sim$800 days post-maximum, while only the 
 initial $\sim$275 days XRT observations were reported 
 by \citet{2022ApJ...930...12H}. The source was X-ray faint, and XRT observed the
 source in photon counting (PC) mode. In this study, we considered all observations 
 with an exposure time greater than 700 seconds to achieve a good
 signal-to-noise. The XRT data reduction followed the standard procedure\footnote{\url{https://www.swift.ac.uk/analysis/xrt/}}. We ran
 \texttt{xrtpipeline} to generate the Level 2 products. We select a circular
 region having a 20-pixel radius around the source and a larger
 circle with a radius of 50 pixels as the background region. No pile-up is
 detected in the source data. We use \texttt{XSELECT} to create the source and
 background spectrum. The \texttt{xrtmkarf} tool with the source spectrum and
 exposure map is used to generate the ancillary response function (ARF) file for
 each observation. We use the response files (RMF) from the CALDB version
20220803\footnote{\url{https://heasarc.gsfc.nasa.gov/docs/heasarc/caldb/data/swift/xrt/index.html}}. All the bad channels are ignored. We group the source
 spectrum with `group min 1' to ensure at least one count per bin and use
 \texttt{cstat} in \texttt{XSPEC} for spectral modelling. The details of the XRT
 observations are tabulated in Table \ref{tab:phot_xrt}.

 \subsection{CHANDRA X-ray Observation}\label{sec:obs:chandra}
 We observed the AT2020ohl with the 
 CHANDRA X-ray space telescope on 2021-04-16, using the High
 Resolution Camera for Imaging (HRC-I) instrument to obtain an image 
 with a high spatial resolution ($\sim0\farcs4$). This observation was also necessary to localize the non-thermal X-ray emission and its association with the possible radio counterparts.
 
\subsection{Other archival data}\label{sec:obs:other}
 The transient occurred at the center of the host. Therefore, it is essential to
 remove the host contribution from the observed flux, to compute the actual flux
 of the transient at different epochs. We used the pre-explosion archival SDSS spectrum of the host center \citep{2013AJ....146...32S},
 which was acquired much before the occurrence of AT2020ohl. We have also used
 the archival photometric data from the SDSS survey 
 to compute the central flux of the host. The procedure is described in
 \S\ref{sec:host}. The ATLAS sky survey \citep{2018PASP..130f4505T} 
 observed the field more than 800 days before the event and
 continued the monitoring afterword. We use ATLAS forced-photometry on the
 host subtracted frames\footnote{\url{https://fallingstar-data.com/forcedphot/}} where there are
 5$\sigma$ detections\footnote{ATLAS usually observes in a quad, i.e. 4 images of
 the same sky area during a given night. To improve the SNR, the fluxes
 of the object obtained from the host subtracted photometry in a given night
 have been stacked. Further, only the 5$\sigma$ detections have been considered
 for this analysis, and corresponding AB magnitudes have been computed.} (see \S\ref{sec:uvopt}).

\section{The Host $-$ NGC6297}\label{sec:host}
 Morphologically, the host is an S0 galaxy\footnotemark[12]. The absence of
 star-forming lines and
 AGN lines in the pre-transient archival SDSS spectrum indicate that the AGN
 activity in this galaxy has been quenched (or at a very low-level). This is 
 also consistent with its X-ray faintness in the pre-transient 
 ROSAT\footnote{\url{https://www.dlr.de/content/en/articles/missions-projects/past-missions/rosat/rosat-mission.html}} X-ray observations. However, 
 analysis of pre-transient $Swift$/XRT observations reveal a faint X-ray luminosity
 $\sim3.4\times10^{41}$ erg$\,s^{-1}$, which points to the existence of
 a low-luminous AGN (LLAGN) at the centre of NGC6297
 (\citealt{2022ApJ...930...12H} and references therein).

 S0 galaxies are found in various environments that indicate multiple pathways of their formation
 (e.g., \citealt{2020MNRAS.498.2372D}). Various possible channels have been proposed
 for their formation $-$ either through gas stripping of spirals inside dense
 galaxy clusters (e.g., \citealt{2000Sci...288.1617Q}) or, by merging of
 disk-dominated galaxies (e.g., \citealt{2015A&A...573A..78Q}) or, due to
 merging of compact ellipticals with their gas-rich irregular companions
 \citep{2018MNRAS.477.2030D}. There are also observational
 evidence suggesting that galaxies with `bars/lenses' make a significant
 contribution to the formation of S0 galaxies, which is difficult to explain with these formation mechanisms
 \citep{2009ApJ...692L..34L}.

 We used the SDSS r-band image to measure the surface brightness properties of NGC6297.
 The left panel of Figure \ref{fig:host} shows the radial profile of the r-band
 intensity of the galaxy in terms of mag/arcsec$^2$ found by 
 fitting concentric isophotes. 
 The outermost isophote is
 at a distance of $\sim$12 kpc from the center of the
 galaxy. The surface brightness of massive galaxies is commonly modelled
 with two components $-$ a Sersic profile for the central bulge and an 
 exponential profile for the disk \citep{2005PASA...22..118G}. We found that along with
 the Sersic profile, two exponential profiles are required to model the
 surface intensity distribution of NGC6297. The
 value of the reduced chi-square ($\chi^2$) for the `Sersic and
 `single-exponential' profile is 12.3,
 while that for the `Sersic and double-exponential' profile is 0.9. This
 implies that the disk of NGC6297 may have two components $-$ an inner disk
 with a steeper intensity profile and an outer disk with shallower intensity
 profile. The presence of an extended disk with multiple components probably  
 supports the
 merging scenario as the origin of the host of AT2020ohl (see \S\ref{sec:discuss}). 

 To estimate the true transient flux (and hence the magnitude) in UVOT
 bands, the flux of the host at different UVOT bands is needed. However as
 the system remained variable over a long timescale in the post-flare epochs, we
 computed synthetic magnitudes of the host-center from the archival data.
 The archival SDSS spectrum (re-calibrated using the $3\farcs5$
 aperture-photometry of the host-center) and the SED produced from the archival
 {\it GALEX} measurements were used. This is shown in the right panel of the Figure
 \ref{fig:host}\footnote{To calculate the central flux of the host, the
 pre-flare images
 of NGC6297 in FUV \& NUV bands (from {\it GALEX} observation), and in NUV-optical
 bands (from UVOT \& SDSS observations) have been used. 
 As discussed in \S\ref{sec:obs:uvopt}, for UVOT data analysis, an aperture of 
 radius of $3\farcs5$ has been used. To calculate the central flux of the host in
 UVOT-filters within an aperture of $3\farcs5$ radius, first we re-calibrated
 the observed flux density of the pre-flare SDSS spectrum of the centre
 (acquired using optical fiber of radius $1\farcs5$) with respect to the magnitudes
 of the host-centre in SDSS $g$ (15.85$\pm$0.05 mag) \& $r$ (15.07$\pm$0.04)
 bands, computed from the pre-flare SDSS images taking an aperture of radius
 $3\farcs5$. Here, we have assumed that the relative flux calibration of the
 SDSS-spectrum is correct. The original (in blue) and the re-calibrated (in
 green) SDSS spectra are shown in the right panel of Figure \ref{fig:host}.
 Further, to construct the SED of the host-centre in UV wavelengths,
 pre-transient observations from {\it GALEX} (in FUV \& NUV bands) and {\it Swift} (in
 $uvw1$, $U$ bands) have been used and computed by fitting a lower-order (order
 3) polynomial to these pre-transient UV data. It is shown with the magenta
 line in the Figure. The magnitudes of the host-center are, respectively 
 20.85$\pm$0.29, 18.89$\pm$0.07, 18.58$\pm$0.20, and 17.56$\pm$0.01 in FUV,
 NUV, $uvw2$, and $U$ bands. After
 constructing the entire re-calibrated UV-optical spectrum of the host-centre
 (i.e., magenta+green lines in the above mentioned Figure), we have computed the
 synthetic magnitudes of the host center (that corresponds to the flux-density
 within an aperture of $3\farcs5$ radius) in $Swift$/UVOT filters (viz.
 $uvw2,~uvm2,~uvw1,~U,~B,~V$)  by using their corresponding response
 curves and following equation 1 of \citet{1986HiA.....7..833K}.}.
 The synthetic magnitudes of the host-centre, in the UVOT {\it uvw2,
 uvm2, uvw1, U, B, \& V} bands, that correspond to flux within an aperture of radius
 $3\farcs5$ are respectively $\sim$18.28, $\sim$18.75, $\sim$17.79, $\sim$17.39,
 $\sim$16.05, and $\sim$15.00 mag\footnote{\citet{2022ApJ...930...12H} computed the
 synthetic-magnitudes of the host-center considering an aperture of radius
 5$''$ and using the synthetic spectrum produced from the code Fitting and
 Assessment of Synthetic Templates (FAST, \citealt{2009ApJ...700..221K}). We
 have reproduced their results by taking an aperture of radius 5$''$. However, in
 the present work, to reduce the host contamination from the beginning, we have
 used $3\farcs5$ aperture to extract the source, and applied aperture-correction.}.
 For the rest of the work, we use these magnitudes as the measures of NUV-optical 
 fluxes of the quiescent galaxy.

\subsection{Distance and Extinction toward NGC6297}\label{sec:DistExt}
 The spectrum of NGC6297 does not have any strong emission lines. The redshift ($z$) of the host computed from its narrow Balmer lines
 is $\sim0.01671$ \citep{2006ApJS..162...38A}. This corresponds to a luminosity
 distance ($D_L$) of $\sim72.9$ Mpc, adopting a standard
 cosmology\footnote{Throughout this paper the cosmology model with $\Omega
_{\rm M} = 0.3$, $\Omega_{\Lambda} = 0.7$ and ${\rm H}_0 = 69.6$
 \kms\,Mpc$^{-1}$ has been assumed \citep{2006PASP..118.1711W}.} corresponding to a distance modulus ($\mu$) $\sim34.3$.

 Estimation of the reddening toward the center of NGC6297 (particularly the
 contribution due to the host) is non-trivial. The Galactic contribution derived from \citet{2011ApJ...737..103S} is 0.0203$\pm$0.0011.
 The inset of the right panel of Figure \ref{fig:host}, shows
 the region around the \Nai\,D$\lambda\lambda5890,5896$ absorption lines in the SDSS
 spectrum of the host. The lines from the Milky-Way and host
 have been marked with cyan and reddish regions. Impression of
 \Nai\,D doublet in the spectrum has been considered as a moderate tracer of
 reddening toward the line of sight over the last couple of decades (e.g.,
 see \citealt{1990A&A...237...79B,1994AJ....107.1022R,2003fthp.conf..200T,
 2012MNRAS.426.1465P}), although recent studies suggest a significant deviation
 from the previously proposed correlations,
 particularly using low-resolution spectra (e.g., see
 \citealt{2013ApJ...779...38P}). The equivalent width (EW) of the \Nai\,D
 doublet due to Milky-Way is $\sim0.1795$\AA\,, that corresponds to a reddening
 of $\sim0.04$ according to \citet{1990A&A...237...79B}, and
 0.023$^{+0.005}_{-0.004}$ according to \citet{2012MNRAS.426.1465P}. Note that
 both measurements are consistent with the above-mentioned value of the Galactic
 reddening. Therefore, the average Galactic reddening
 along the direction of the transient is $\approx0.028$. \Nai\,D absorption
 dip of the host is more prominent (although the components of the doublet are
 unresolved) than that of
 Milky-Way. The EW of this unresolved line is 2.63$\pm$0.22\AA\, (the error in
 EW is estimated using Equation 6 of \citealt{2006AN....327..862V}). This
 corresponds to a reddening from the host galaxy 0.66$\pm$0.05 according to
 \citet{1990A&A...237...79B} and to \citet{2012MNRAS.426.1465P}
 corresponding value of reddening is $\sim$16, which is essentially infeasible.

 Therefore, in this work, we presume that the contribution of the host in
 reddening toward the center of NGC6297 is negligibly small. 
 This is mainly supported by the following attributes $-$ (a) the
 transient remains highly luminous in the NUV for a long time (see
 \S\ref{sec:evol}), (b) the host is a S0 galaxy, therefore the star formation in its centre may have been quenched,
 causing a very less abundance of dust along the center of the host. 
 Therefore, considering only
 the Galactic absorption, the total reddening toward the transient is
 \ebv\,$\approx0.028$.
 This implies the total extinction toward AT2020ohl in V-band is
 $A_V\approx$0.09 mag, adopting the uniform value of total-to-selective
 extinction ratio ($R_V) =$ 3.1 of the Milky-Way.

\section{Evolution of AT2020ohl}\label{sec:evol}
\begin{figure}
\includegraphics[width=8.5cm,angle=0]{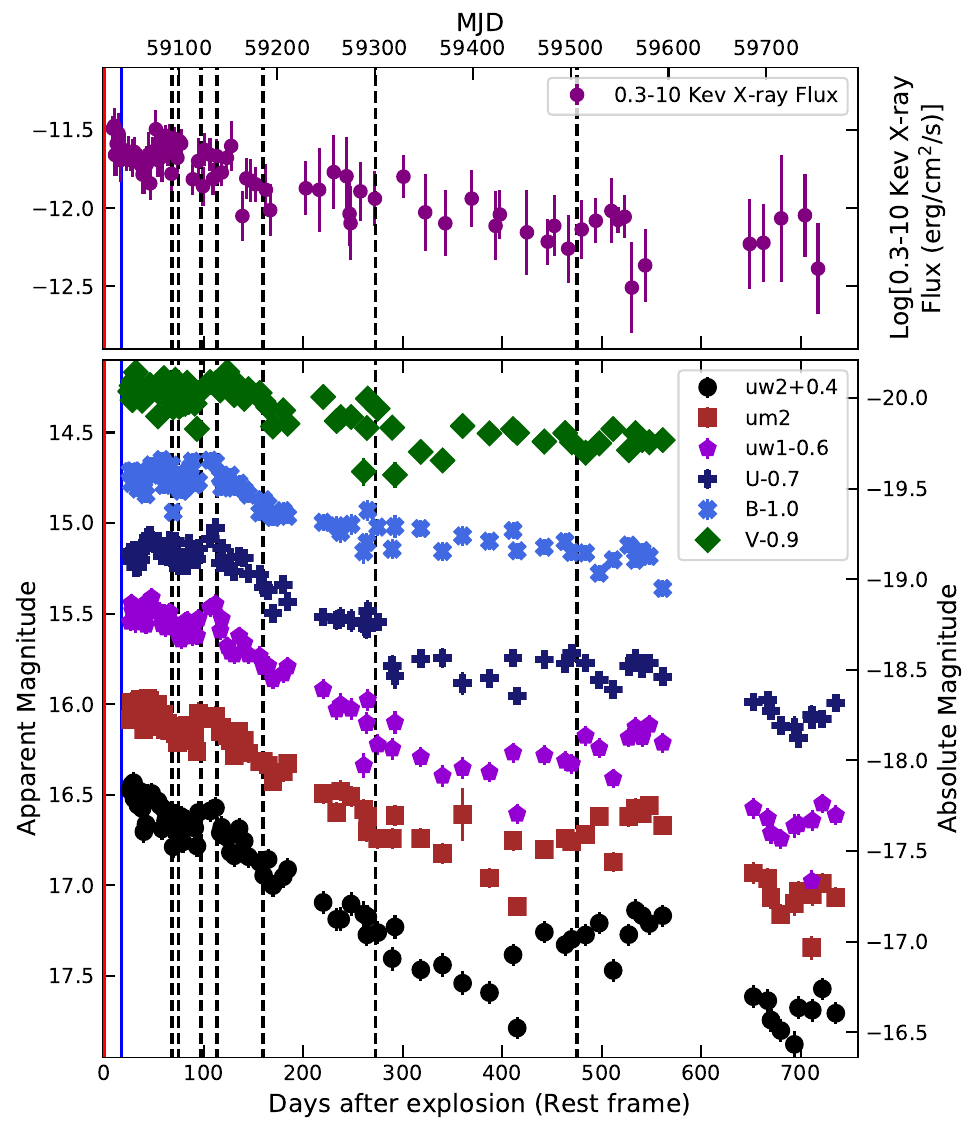}
\caption{The X-ray, UV, and optical lightcurves of AT2020ohl. The
 {\bf Lower Panel} shows the UV-optical lightcurve of the transient. All the
 observed magnitudes of the transient have been obtained by subtracting the host
 central flux from the measured fluxes of the central region at different epochs
 as detailed in \S\ref{sec:host}. Variations in magnitudes in small and large
 timescales are noticeable. 
 {\bf Upper Panel} shows the corresponding variations in the X-ray fluxes.
 The red and blue vertical lines respectively mark
 the last non-detection and first detection reported by ASASSN. Note that the
 last non-detection was noticed $\sim$2 days before the epoch when the first rise
 was observed by TESS (\S\ref{sec:obs}) The black dashed
 vertical lines mark the epochs when HCT spectroscopic observations were
 performed.}
\label{fig:LC}
\end{figure}

 Using the high-cadence pre-transient data from TESS \citet{2022ApJ...930...12H}
 found a smooth rise in flux from the pre-transient to the transient state,
 following a power-law with temporal index $\sim$1.05. In this work, we will mainly
 focus on the post-maximum evolution of AT2020ohl. Figure \ref{fig:LC} shows the
 X-ray, UV and optical light curves of the transient. The red and blue vertical
 lines mark the onset of the event (from TESS observation) and the first detection reported
 by the ASAS-SN survey (see \S\ref{sec:obs}). 

\subsection{UV-optical lightcurve}\label{sec:uvopt}

\begin{figure*}
\centering
\includegraphics[width=8cm,angle=0]{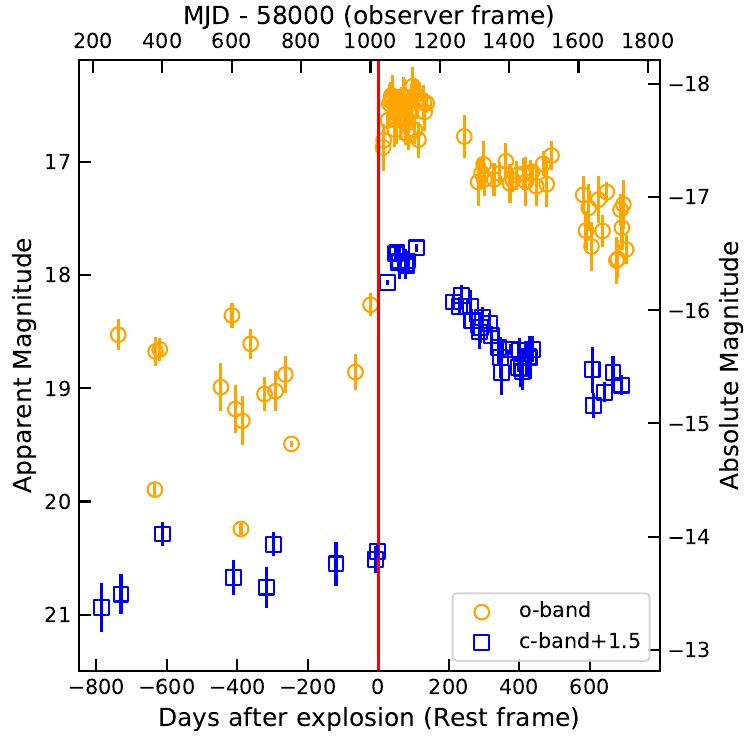}
\hskip 0.5cm
\includegraphics[width=8cm,angle=0]{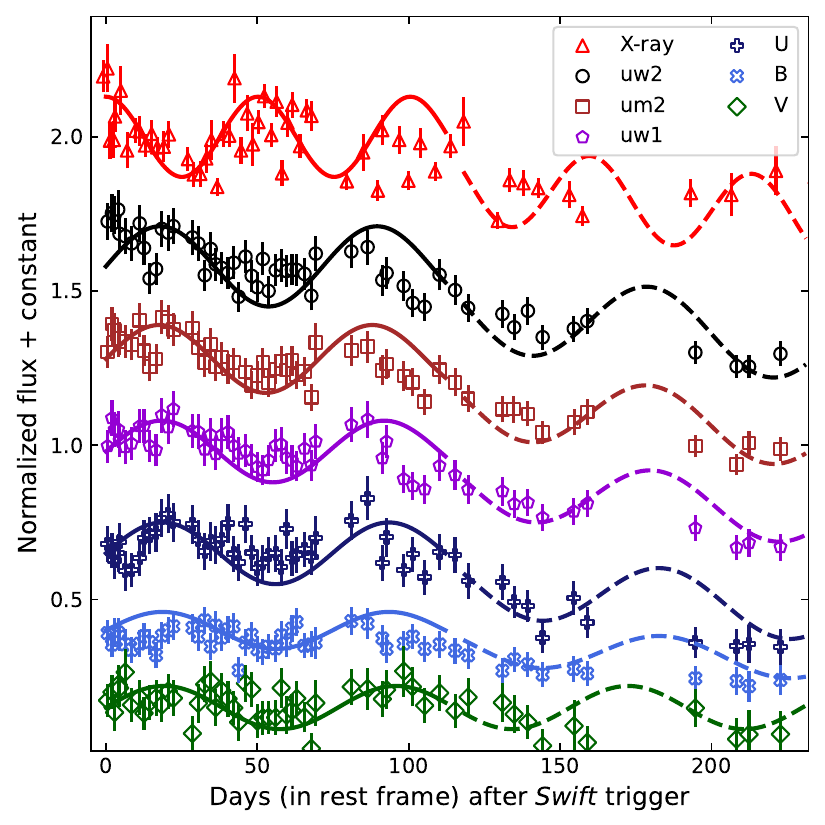}
\caption{The X-ray, UV, and optical emission from AT2020ohl. 
 {\bf Left Panel} shows the lightcurves of AT2020ohl in ATLAS o and c-bands. 
 {\bf Right Panel} shows only the early part (post-maximum) of the lightcurve
 observed from {\it Swift}.
 Significant variations in X-ray, and UV-optical light curves of the transient are noticeable. The solid and dashed lines over-plotted with the data highlight the variation of flux (qualitatively) at the early and late phases after the maximum respectively.}
\label{fig:LCperiod}
\end{figure*}

\begin{figure*}
\centering
\includegraphics[width=7.5cm,angle=0]{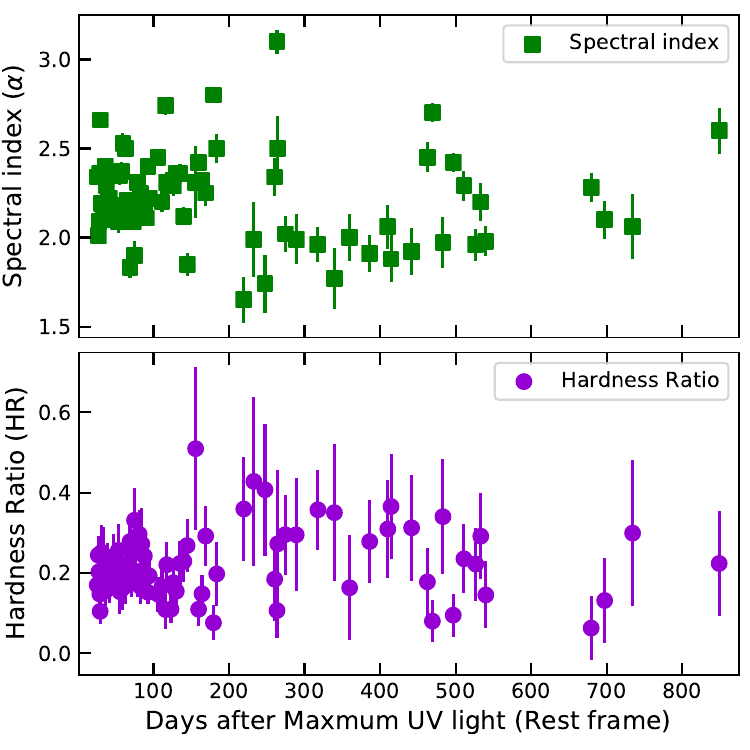}
\hskip 0.5cm
\includegraphics[width=9.5cm,angle=0]{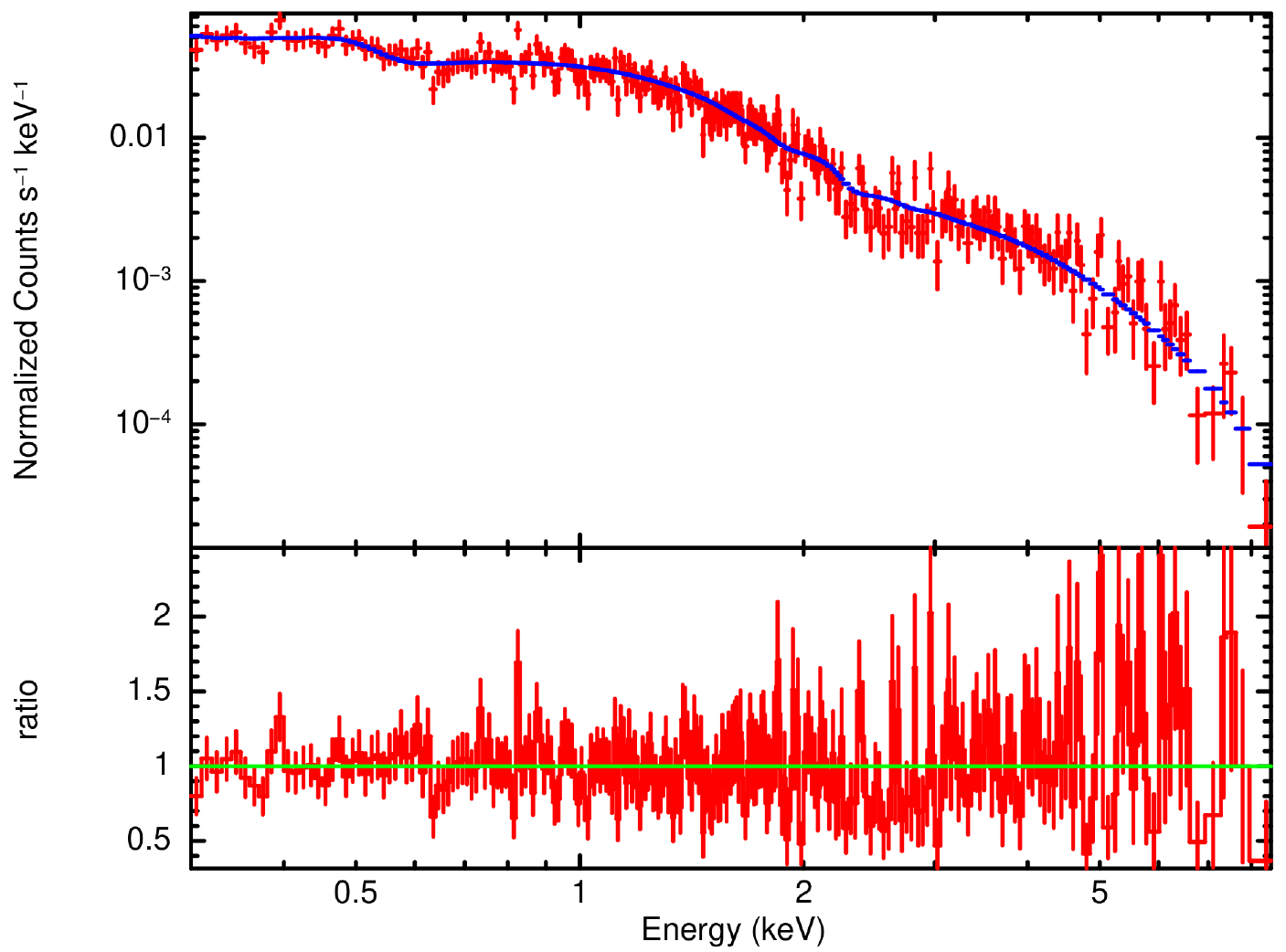}
\caption{The {\bf Left panel} shows the variation of hadness ratio (HR) and
 spectral index ($\alpha$) of the X-ray emission from the source during its
 evolution. The {\bf Right panel} shows the average X-ray spectrum of the
 transient during its evolution in first 100 days. See text for the detail.}
\label{fig:xspec_HR}
\end{figure*}

 Flat/quasi-flat lightcurves are prominent during the first 120 days in all
 UVOT-bands, followed by a
 gradual decay until $\sim+400$ days in the rest frame of the transient. Assuming
 a power-law decay of the flux with time ($f\propto t^{-\alpha}$), two distinct
 decay profiles were seen before and after +120 day in each lightcurve.
 Before +120 day, the temporal index ($\alpha_1$) remained almost constant in all
 UVOT bands ($0.07\pm0.01$ in $uvw2$, to $0.01\pm0.01$ in $V$-band). Between
 $120-400$ days, a steeper decay (with index $\alpha_2$) is observed. It is more prominent in  the NUV bands, with a larger value of $\alpha_2$
 (0.41$\pm$0.02), while shallower in $U$-band (roughly 0.3) and even
 shallower in $B$ \& $V$-bands (roughly 0.1). Beyond +400 day, a shallow rise is seen for all of the 
 NUV bands for the next 150 days followed by a decay until +700 day.
In the $B$ and $V$ bands, the corresponding variations are almost negligible, although a decay of the $U$-band flux beyond +550 day is also noticeable.

 The left panel of Figure \ref{fig:LCperiod} shows the ATLAS forced-photometry
 of the center of NGC6297, performed on the host subtracted images in the ATLAS
 Orange (o) \& Cyan (c)-bands and covering a timespan from $\sim-800$d to +800d
 with respect to the onset of the event (i.e., t$_0$). Prior to that, fluctuations around the median of 19 mag are noticeable in both ATLAS bands. However, from
 this data, it would be exaggerated to say that those fluctuations are
 precursors of AT2020ohl. Nevertheless, it is noteworthy that on MJD =
 59000.4 (i.e., 22 days before the onset of the event in the source frame), the
 object became brighter than the median by 0.7 mag in
 the o-band. This flux enhancement is within 2$\sigma$ from the median-value (where
 standard deviation for the pre-event o-band measurements is $\sim0.5$
 mag). The trend in the enhancement of flux in the o-band during that period is more
 prominent in 3$\sigma$ clipping analysis and, to some extent, is also noticeable
 in TESS observations (see Figure 4 of \citealt{2022ApJ...930...12H}). This
 may be a precursor of AT2020ohl. However, a rise in the c-band was not observed in
 either of the 5$\sigma$ or the 3$\sigma$ clipped photometry. So, if
 it is indeed a precursor of AT2020ohl, the corresponding spectrum must be very
 reddish. 
 
 On the other hand, beyond +400 day a little rise in both o-band and c-band
 fluxes can be seen. As the host galaxy is faint in NUV and
 $U$-bands, a flux enhancement during that period has also been found from the
 analysis of {\it Swift} data. The enhancement in NUV is noticeably higher than
 other optical bands.
 
 A close inspection of the first 120 days of photometric evolution after the
 peak reveals significant variation in the UV-optical light curves (right panel of Figure \ref{fig:LCperiod}). The solid lines with periodicity $\sim$11 days are over-plotted with the data. 
 Similar behaviour has also been noticed in X-ray
 lightcurve (with a periodicity of $\sim$8 days) during this timespan. 
 However, due to the lack of high-cadence data with regular intervals, we cannot perform any quantitative analysis to conclusively establish the periodic variation in the light curves. 
 
\subsection{X-ray lightcurve and spectrum}\label{sec:x-ray}
 The upper panel of Figure \ref{fig:LC} shows the X-ray lightcurve. Like 
 UVOT data, all of the orbit data of a given {\it Swift}/XRT observation were stacked 
 to improve the SNR. 
 We fit the individual, as well as the time-averaged X-ray spectra 
 with the \texttt{ztbabs*powerlaw} model with a line of sight Hydrogen column density $n_H=1.3\times10^{20} cm^{-2}$. The photon index of the time-averaged X-ray spectrum is $\sim2.1$ (see the right
 panel of Figure \ref{fig:xspec_HR}).
 From the spectral modeling, we estimated the hardness ratio (HR$-$defined as the ratio of
 2-10 keV and 0.3-2 keV flux) and the spectral index for all the {\it Swift}/XRT observations.  The HR (see lower plot 
 of the left panel of Figure \ref{fig:xspec_HR}) and spectral index ( upper plot 
 of the left panel of Figure \ref{fig:xspec_HR}) remain roughly constant at
 around 0.2 and 2, respectively. This indicates the soft nature of the X-ray emission. These results are consistent with the
 findings of \citet{2022ApJ...930...12H}. 
The X-ray emission from AT2020ohl was completely power-law dominated and plausibly 
generated through non-thermal processes, while as described in \S\ref{sec:dis_uvop} 
and \S\ref{sec:dis_comb}) the UV-optical photons were produced due to black body 
emission. This is also evident from the uncorrelated nature of X-ray and UV-optical 
light curves.

\subsection{Optical spectra}\label{sec:optspec}
\begin{figure}
\includegraphics[width=8.5cm,angle=0]{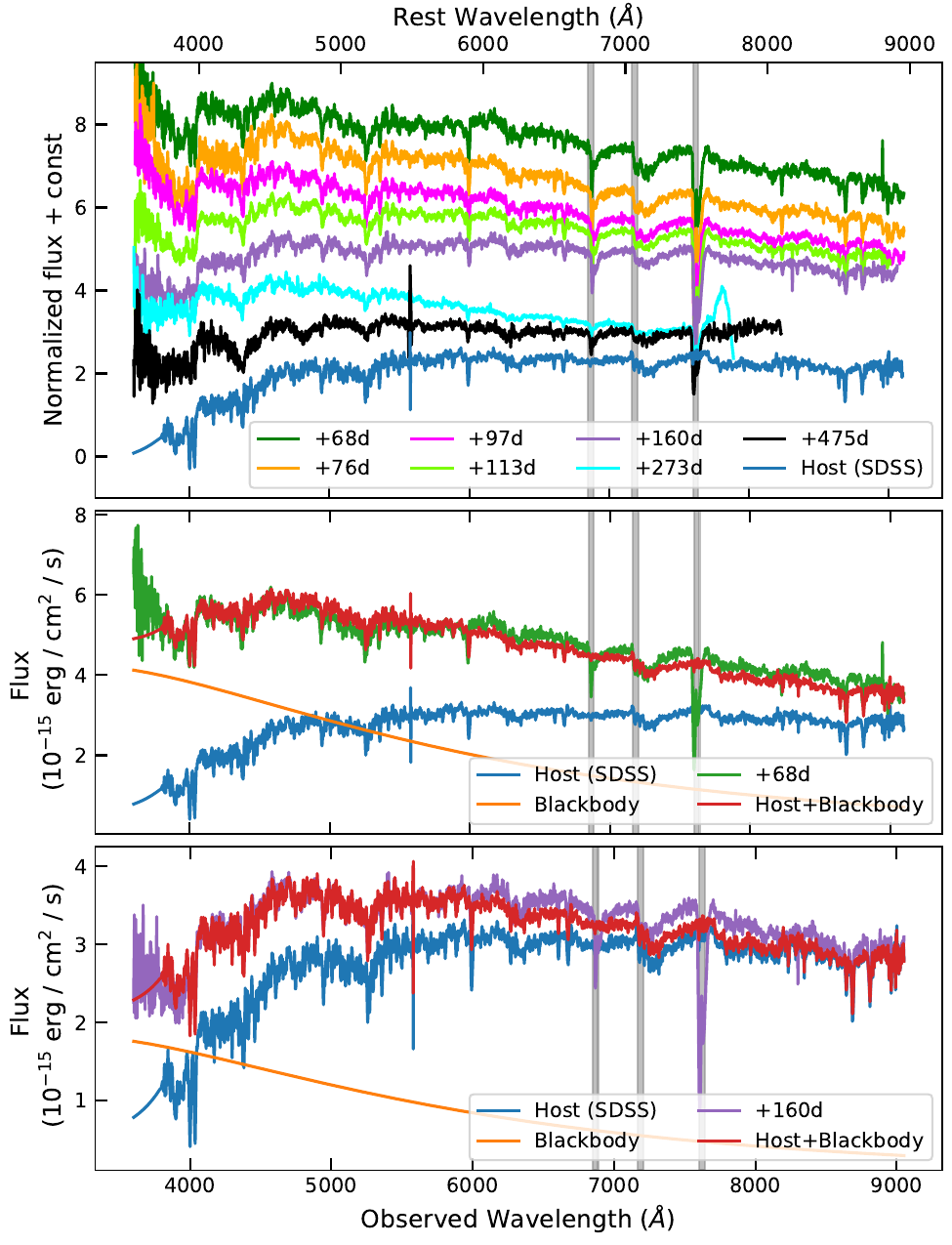}
\caption{Spectroscopic evolution of AT2020ohl. The
 {\bf Upper Panel} shows the spectra of the transient observed at +68d, +76d,
 +97d, +113d, +160d, +273d, and +475d. The SDSS spectrum of the host
 galaxy has also been shown. Over the timescale of $\sim$500 days after the 
 onset of the event, no strong line due to a TDE-like event has been observed.
 Narrow vertical grey regions mark the positions of teluric lines.
 {\bf Middle and Bottom Panels} demonstrate the fitting of the Blackbody spectrum at
 two different epochs (+68d \& +160d). The addition of the Blackbody spectrum
 (characterized by a single temperature `T') emitted by a spherical region of
 radius `R', with the host spectrum, can reproduce the transient spectrum at that
 epoch. For the +68d optical spectrum, the temperature and radius are, respectively
 T$_{68} = 8966\pm50$K \& R$_{68}\sim3.1\times10^{15}$cm, whereas the +160d
 spectrum can be reproduced with a blackbody surface at T$_{160} = 9100\pm115$K
 \& R$_{160}\sim1.9\times10^{15}$cm. These show there is not much evolution of
 the transient between these two epochs. The implication of this analysis has
 been discussed in \S\ref{sec:discuss}.} 
\label{fig:spec}
\end{figure}
 
 The spectral evolution of the transient in optical wavelength is shown
 in Figure \ref{fig:spec} along with the SDSS pre-transient spectrum of the host. From the upper panel of Figure \ref{fig:spec}, it
 is clear that during its entire evolution (which corresponds to a timescale of
 $\sim500$ days post disruption in the rest frame), no new line was produced due to the transient activity. This may rule out the possibility of a TDE or SN as the origin of AT2020ohl (see
 \S\ref{sec:discuss}). As shown in the
 middle and bottom panels, the +68d spectrum of the transient(+host) can be
 reproduced by adding a black body component to the host spectrum, where the
 black body emission is produced by a photosphere with a radius of 
 R$_{68}\sim3.1\times10^{15}$cm and having an effective temperature
 T$_{68} = 8966\pm50$K.
 Similarly, the +160d spectrum of the transient(+host) can be reproduced by
 adding a black body component with R$_{160}\sim1.9\times10^{15}$cm 
 and T$_{160} = 9100\pm115$K to the host spectrum\footnote{Noteworthy, 
 the resolution of SDSS spectrum is comparable to that of the 
 observed spectra of the transient. To perform the fitting, we first 
 calculated the flux values of every observed spectrum at the wavelength 
 bins of the SDSS spectrum using cubic spline interpolation. Further, we 
 added a black body component with the SDSS spectrum and fitted it to 
 individual observed spectrum using \texttt{LMFIT} package by 
 implementing $\chi^2$ minimization technique while 
 varying the radius and temperature of the black body.}. Therefore, the 
 optical
 emission produced by the transient had a black body origin, and the nature of the
 source did not change appreciably during the first 200 days. Later the emitter 
 became cooler, and slowly the transient intensity started to decline (see the 
 bolometric lightcurve in \S\ref{sec:discuss}).

\subsection{Evolution in Radio-band $-$ connection with X-ray}\label{sec:radioevol}
\begin{figure}
\includegraphics[width=8.5cm,angle=0]{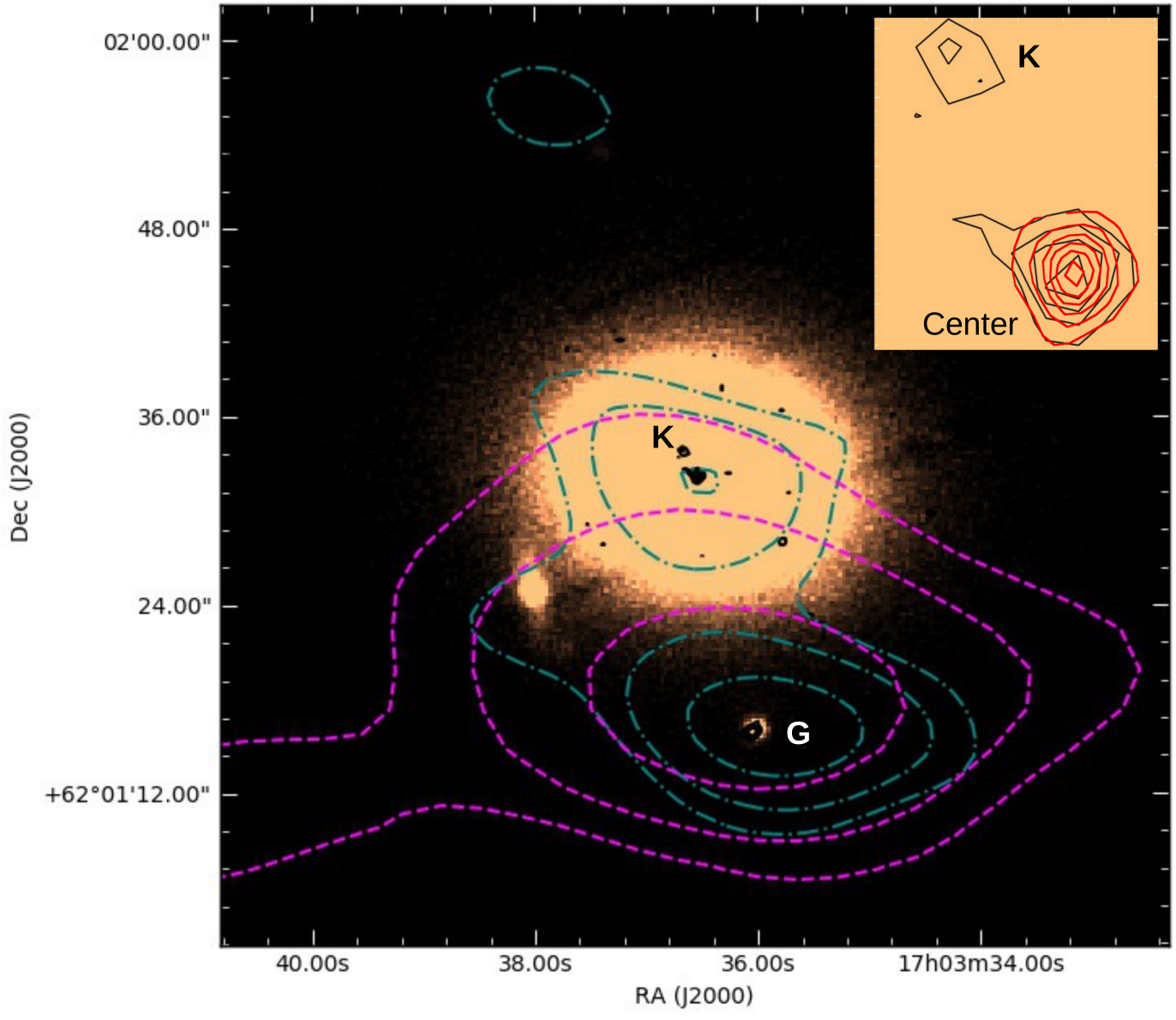}
\caption{This figure shows the JVLA radio and CHANDRA X-ray counterparts of AT2020ohl, 
superposed on the archival optical image of the host (the bright big galaxy at the 
centre) taken from the PanSTARR survey (North is up and East is in the left direction). 
The JVLA 
X-band (B-array) contours (black) show the compact radio component distinguishing the 
centre (O), nearby knot (K) in the north-east, and the nearby compact radio source `G'. 
The central region has been magnified in the inset where CHANDRA X-ray counterpart (in 
red) has also been shown. The green dot-dashed contours show the JVLA X-band (D-array) 
observations, while the magenta dashed contours show the JVLA C-band (D-array) 
observations. The X-band B array contours have been plotted at 3$\sigma$
 [1, 1.5, 2, 2.5], where the rms noise is $\sigma$ = 3 $\mu$Jy beam$^{-1}$. For the
 X-band D array also, the contours have been plotted at 3$\sigma$ [1, 2, 4], where the 
 rms noise is $\sigma$ = 6$\mu$Jy beam$^{-1}$. For the C-band D 
 array contours have been plotted at 3$\sigma$ [1, 2, 4], where
 the rms noise is $\sigma$ = 5 $\mu$Jy beam$^{-1}$.
}
\label{fig:OptXrayRadio}
\end{figure}

 As described in \S\ref{sec:obs:radio}, the transient was detected in the JVLA X and C
 bands, at two epochs separated by 7 months. The radio flux contours
 superimposed on the SDSS optical image for these two observations are shown in
 Figure \ref{fig:OptXrayRadio}. The X-band observation in the 
 higher resolution under B-array configuration (synthesized beamwidth
$\theta=0\farcs6$), resolves the central emission (O) from the nearby
 radio knot `K' to the $1\farcs5$ north-east, as well as the nearby 
 galaxy with compact radio emission $\sim$16\arcsec to the south and marked as `G'. 
 The integrated X-band flux density of the central core on +111 day is 33.5$\pm$3.7 
 $\mu$Jy. The corresponding measurement for the knot (K) is 18.4$\pm$3.7 $\mu$Jy.

 Although the event was X-ray luminous, the transient can not be well localized due to the low resolution of the {\it Swift}/XRT (18\arcsec at 1.5 keV). 
 To localize the X-ray transient, we observed it with CHANDRA/HRC-I.
 The upper-right inset shows the central region, and the centroid of the X-ray source, is consistent with the location of the central radio counterpart, confirming the nuclear-origin 
 of AT2020ohl, and that the nearby radio-knot is not associated with the present transient activity (see \S\ref{sec:discuss}).  

The lower resolution D-array X-band images ($\theta=7\farcs2$) could resolve the emission of 
 NGC6297 (O + K) from the nearby galaxy G (see the green dot-dashed contour), 
 but cannot 
the emission from the center of the host. The integrated X-ray flux density in
 the D-array data is 65.08$\pm$14 $\mu$Jy. Since the nearby knot is not
 associated with the transient activity, its X-band luminosity is expected to be 
 constant 
 and we can conclude the transient X-band flux had been increased to $\sim47\pm14.7$
 $\mu$Jy by +313 day. This is nearly 1.4 times higher than the X-band flux of the 
 transient
 at +111 day, although the two values are consistent given 
 their errors. This shows that over 200 days of evolution, there
 was no significant change in the X-band luminosity of AT2020ohl.
 On the other hand, the D-array C-band observation ($\theta=12$\arcsec) on
 +318 day could resolve none of the three sources therefore gives only the upper limit 
 on the flux of the transient
 (see the magenta dashed contour). The integrated flux density of the
 entire C-band contour shown in the Figure is 132.78$\pm$20 $\mu$Jy. It is
 also noticeable from this image that the centroid of this contour coincides
 with the position of the nearby galaxy (G). It implies that G mostly dominates
 the radio flux. The peak 
 flux density of this contour is 101.41$\pm$10 $\mu$Jy/beam. Therefore the 
 upper limit of the central (i.e., transient) C-band flux of NGC6297 is
 roughly 30 $\mu$Jy on +318 day (we consider it as the upper limit as this
 flux is a combination of transient flux and that of the radio-knot in C-band).
 
\begin{figure}
\includegraphics[width=8cm]{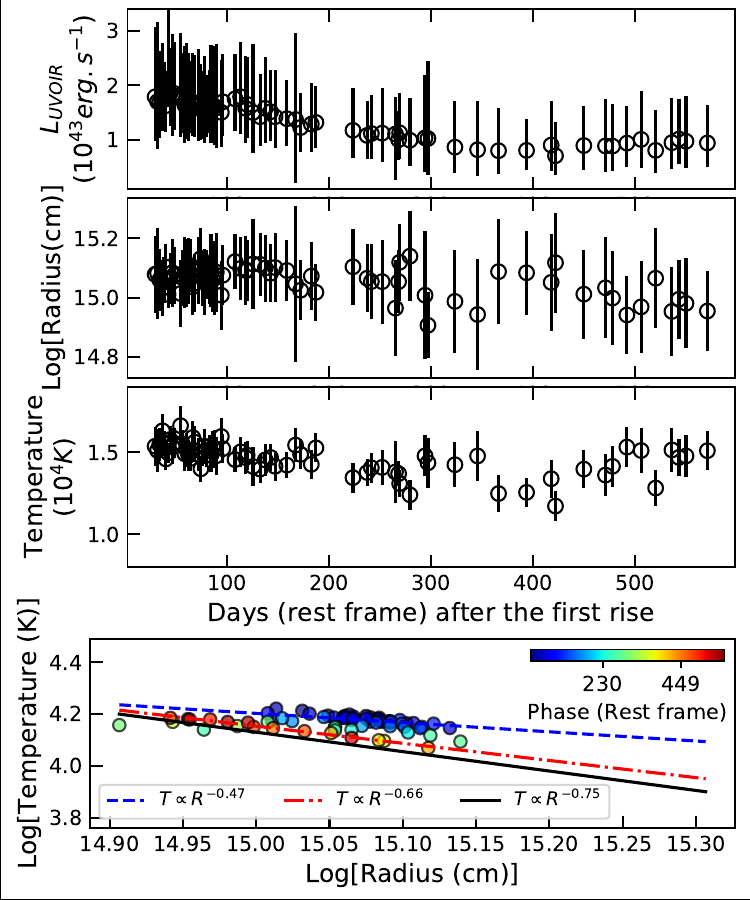}
\caption{The UVOIR bolometric lightcurve of AT2020ohl.
 From top, 
 the first, second and third panel respectively describes the temporal evolution
 of bolometric luminsity (L$_{UVOIR}$), Radius of the emitting surface, and its
 Temperature. The lower most panel shows the observed variation of tempearture 
 with radius. The color-bar indicates the phase in rest frame. The blue dash 
 line is fit to the data having phase value $\textless200$ days in rest frame
 showing T$\propto$R$^{-0.47}$. The red dot-dash line is fit to the data having
 phase value $\textgreater400$ days in rest frame showing T$\propto$R$^{-0.66}$.
 The black solid line shows the classical temterature-radius relation for an
 accretion disk (T$\propto$R$^{-3/4}$).} 
\label{fig:bol}
\end{figure}

\section{Discussion and Conclusions}\label{sec:discuss}
 \citet{2022ApJ...930...12H} pointed out that AT2020ohl/ASASSN-20hx simultaneously exhibited the characteristics of TDEs and AGN. Substantial emission in NUV, the
 smooth rise of the TESS light curve in TESS-band and the overall UV/optical evolution
 make the object comparable to TDE-like events. On the other
 hand, the linear rise of the flux,
 the absence of emission lines (also described in
 \S\ref{sec:optspec}) differentiate AT2020ohl from canonical TDEs. Moreover, the
 non-thermal origin of the X-ray and the location of NGC6297 in the NIR color-color diagram
 is evidence for a possible AGN-origin of this event (also see \citealt{2018ApJ...852...37A}), although strong AGN
 lines were absent until the +483d spectrum (see Figure \ref{fig:spec},
 and also \citealt{2022ApJ...930...12H}). In this work, we
 have revisited this problem to understand the origin of AT2020ohl using
 multi-wavelength data. This is also to explore the probable mechanisms behind some of the ambiguous nuclear transients.

\subsection{The nature of NUV \& optical radiation}\label{sec:dis_uvop}
\begin{figure*}
\centering
\includegraphics[width=8.5cm,angle=0]{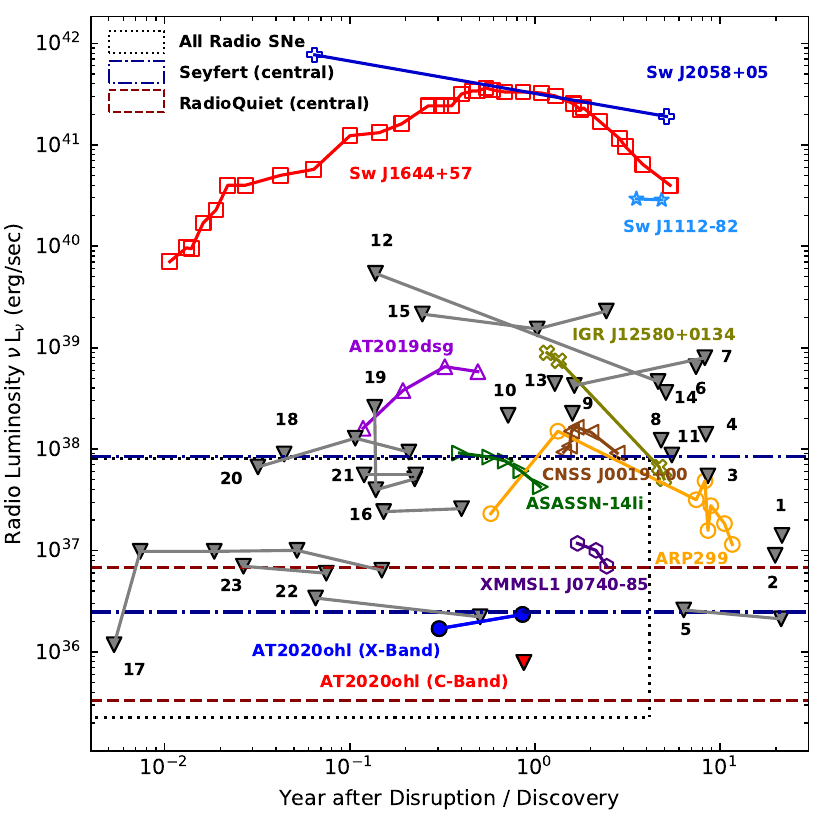}
\hskip 0.5cm
\includegraphics[width=8.5cm,angle=0]{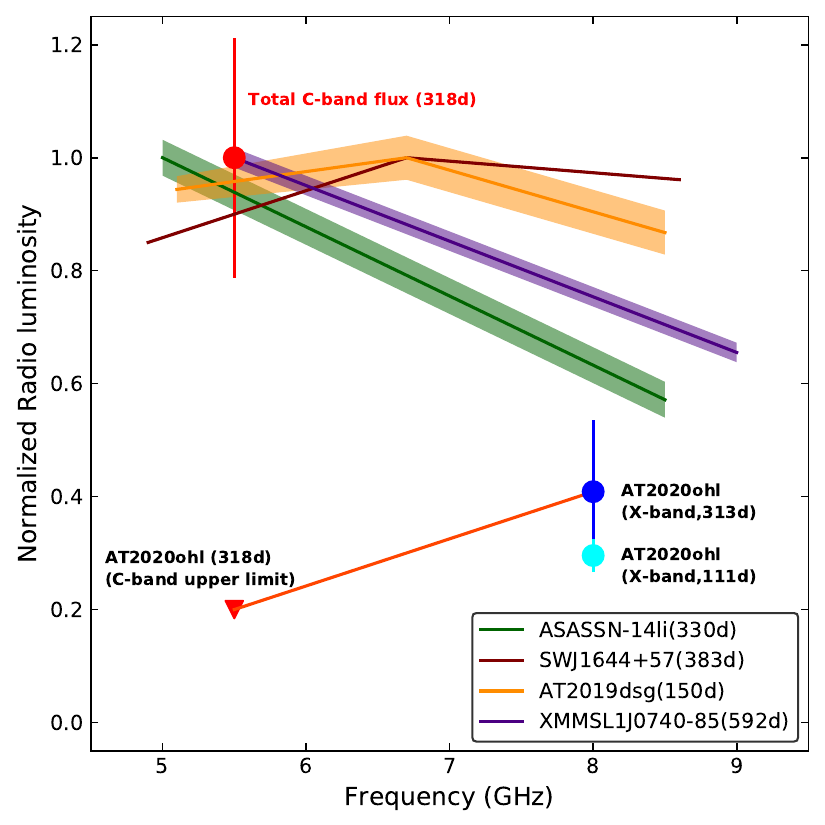}
\caption{Evolution of AT2020ohl in radio-band. {\bf Left Panel} shows the light
 curve and its comparison with other energetic sources, viz. radio counterparts
 of SNe (region marked with dotted line), emission from the central part
 of Syefert galaxy (region marked with dot-dashed line), and central radio
 luminosity of Radio-Quiet AGNs (region marked with dashed line), and
 radio-TDEs (other points). The figure has been adopted from
 \citet{2020SSRv..216...81A} and further modified. The downward gray triangles
 are showing upper limits of the fluxes of several TDEs which have been labeled
 with numbers as per first column of Table 2 of \citet{2020SSRv..216...81A}.
 {\bf Right Panel} shows the comparison of radio spectra of AT2020ohl with the
 radio-spectra of some radio luminous TDEs at comparable (and late) epochs, viz.
 SWJ1644+57 \citep{2013ApJ...767..152Z}, ASASSN-14li, XMMSL1 J0740-85
 \citep{2016ApJ...819L..25A, 2017ApJ...837..153A}, and AT2019dsg 
 \citep{2021MNRAS.504..792C}.}
\label{fig:radio}
\end{figure*}

 Although the rise of AT2020ohl was smooth (a monotonic function of time,
 \citealt{2022ApJ...930...12H}), its post-maximum evolution in different
 UV-optical bands were significantly different from those of canonical stellar
 disruptions (see \S\ref{sec:uvopt}). 
 The {\it Swift} NUV/optical lightcurves showed quasi-periodicity for nearly 
 120 days after the maximum (lower panel of Figure \ref{fig:LC}). This
 indicates that the origin of these photons is roughly the same (although not 
 precisely as discussed below).

 The upper panel of Figure \ref{fig:bol} shows the UV/optical bolometric lightcurve
 of the object along with the variation of the photospheric radius and temperature of 
 the transient. To compute the bolometric lightcurve, we used the host-subtracted 
 photometry of the transient obtained from the UVOT and ATLAS
 observations. The lightcurve was computed at the epochs of the Swift
 observations, and the observations from ATLAS were linearly interpolated to the {\it 
 Swift} epochs. The observed SED of every epoch can be explained with a single black 
 body. The temporal variation of black body radius and temperature have been shown 
 in the second and third panels of the Figure \ref{fig:bol}.
 Although the peak luminosity is consistent with canonical SNe, or TDEs, the post-peak 
 decline rate is extremely shallow $-$ maintaining a power
 law with temporal index $\sim-$0.07 during the first 120 days, and thereafter a
 steeper index of $\sim-$0.3 has been observed (where negative sign indicates a decrease in flux with time, \S\ref{sec:uvopt}). Among nuclear
 transients, similar slow bolometric evolution has been observed in a few cases
 like ASASSN-17cv, ASASSN-18el \citep{2019NatAs...3..242T, 2019ApJ...883...94T},
 and ASASSN-18jd \citep{2020MNRAS.494.2538N}. All these are  
 ambiguous transients $-$ either changing-look AGN (ASASSN-18el) or rejuvenated
 SMBH (ASASSN-17cv) or some unknown SMBH-driven transient in an AGN system
 (ASASSN-18jd).

 The temperature obtained from the spectra are relatively lower than that computed from the 
 SED-fitting (however, it is not an
 order of magnitude higher). This is mainly because SED contains
 spectral information over a larger wavelength range, and the 
 photometric flux calibrations are better than spectroscopic flux 
 calibrations. The higher value of inferred temperature, while 
 using the UV-optical photometry was noticed previously as well in 
 nuclear transients like TDEs (e.g., ASASSN-14ae, 
 \citealt{2014MNRAS.445.3263H}). One probable reason may be, that 
 although we 
 are assuming the origin of UV-optical photons is the same, actually UV 
 photons are coming from the relatively inner hotter part of the disk, 
 while the optical photons are generated at the relatively outer cooler 
 part of the disk. The temperatures computed from the optical spectra are
 therefore lower than that calculated from the UV dominated SED. Noteworthy, our last 
 optical spectrum was taken on +475 day. So, at late-epoch, when UV is relatively less 
 dominant, the photometric results become more consistent with spectroscopic results. 
 However, beyond +475 day, we could only compute black body temperature and radius from 
 photometric measurements. Therefore, with the existing data set, it is not possible to 
 compare the results from photometry and spectroscopy beyond +475 day.
 Important information about the emitter of UV-optical radiation can be drawn from the correlation of temperature with radius 
 shown in the fourth panel of the Figure. 
 In the early evolutionary phases ($\textless200$ days in rest frame), the radius (R) and 
 temperature (T) follow the relation T$\propto$R$^{-0.47}$ (blue dash line), while in the later
 phases ($\textgreater400$ days in rest frame), they follow the relation
 T$\propto$R$^{-0.66}$ (red dot-dash line). The black solid line shows the
 radius-temperature relation for a standard accretion disk
 (T$\propto$R$^{-3/4}$, \citealt{1973A&A....24..337S}). 
 The transition of the radius-temperature relation from the shallower slope at the early epoch 
 towards the standard accretion disk scenario at the late epoch essentially demonstrates that
 UV-optical emitting region has an accretion-disk like structure, which was
 initially evolving dynamically (just after the event), settled down with time, and approached
 to a steady accretion disk in due course of time.

\begin{figure*}
\centering
\includegraphics[width=17cm,angle=0]{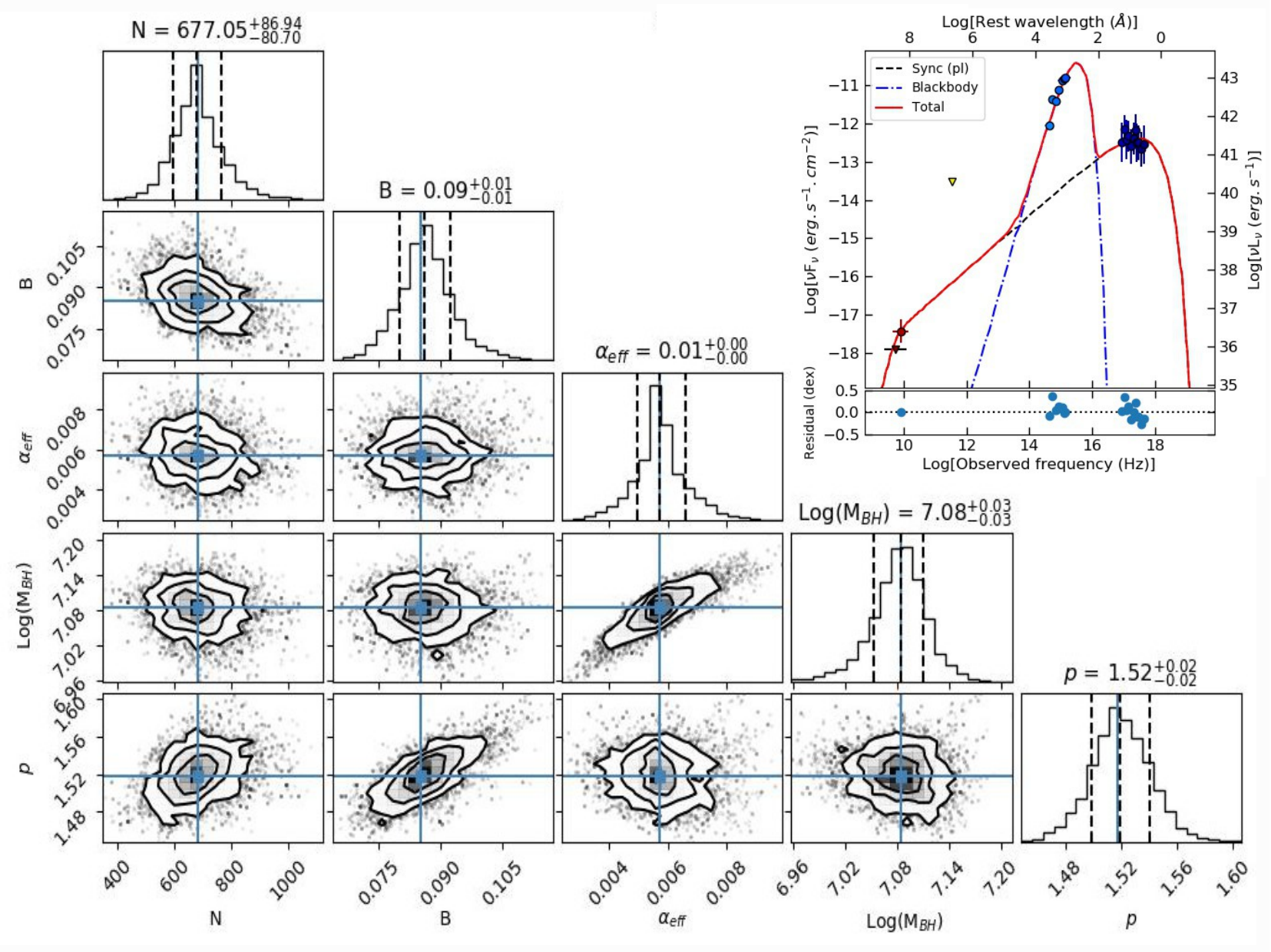}
\caption{The broad-band SED modelling of AT2020ohl. X-ray and Radio are produced
 from the synchrotron process, considering the single powerlaw model. However,
 the NUV-optical emission is mainly from the Blackbody radiation from the disk.
 The upper right inset shows the observed data points and different components
 of the model, along with their sum (in red) and residual. The corner plot shows
 the posterior distribution of 5 variables in the MCMC analysis.
}
\label{fig:SEDmod}
\end{figure*}

\subsection{Origin of X-ray \& Radio photons}\label{sec:dis_xrad}
 Thanks to the quasi-simultaneous, high-resolution X-ray and radio observations, it is clear 
 that, the X-ray and radio emissions originate from the same loation. Figure \ref{fig:radio} 
 shows the nature of radio emission from the central component shown in 
 Figure \ref{fig:OptXrayRadio}. The Left panel of Figure \ref{fig:radio}
 shows the post-disruption temporal evolution of the X-band radio luminosity between days 
 $\sim111$ and $\sim313$. It also shows the upper limit of the C-band radio luminosity at 313 
 day and compares it to the 5 GHz radio luminosities of other transients (adopted from
 \citet{2020SSRv..216...81A}) like TDEs,
 radio-luminous SNe, and the central radio luminosities of Seyfert galaxies and
 radio-quiet AGNs.
 Clearly, AT2020ohl is several orders of magnitude fainter than other radio transients. As 
 shown in the Figure, the X-band luminosity of AT2020ohl increased only by 1.3 order in 200
 days. This evolutionary rate is much slower than that of the other radio-TDEs shown in the 
 Figure. At +313 days post-disruption, it was 10 times dimmer than the central
 radio-luminosity of low-luminous Seyfert galaxies and low-luminous radio-quiet AGNs. At a 
 similar phase of evolution, its radio luminosity is comparable only with some of the
 low-luminosity radio SNe. However, it had no other signature that suggests it could be a 
 supernova. Given its spectral properties at other wavelengths, particularly in the 
 optical and NUV, we rule out the possibility of the association of
 AT2020ohl with any stellar explosion or disruption phenomenon $-$ it is neither a supernova
 nor a tidal disruption event. Nevertheless, the gradual increase of its high-frequency radio 
 luminosity marks the possible slow emergence of an outflow from the center of NGC6297.

The normalized radio spectra of AT2020ohl are compared with some of the radio-luminous TDEs in the right panel of Figure \ref{fig:radio}. Here, cyan and blue
 points are the X-band luminosities of the transient at +111 day and +313 day, respectively. The red-filled circle corresponds to the total integrated flux in C-band, and the downward triangle represents the upper limit of the C-band transient flux (see \S\ref{sec:radioevol}). 
 The orange straight line (between the C-band upper limit and the X-band detection) shows the radio SED of the transient at $\sim$300 days after the disruption. The positive value
 of the spectral index (assuming $F\propto\nu^\beta$) shows it is in the optically-thick phase 
 at this stage of its evolution. The radio SED of AT2020ohl is also compared with 
 the SEDs of the radio luminous TDEs observed at comparable epochs $-$
 SWJ164449.3+573451 (at +383d, \citealt{2013ApJ...767..152Z}), XMMSL1J0740‑85
 (+592d, \citealt{2017ApJ...837..153A}), ASASSN-14li (+330d,
 \citealt{2016ApJ...819L..25A}), and AT2019dsg (+150d,
 \citealt{2021MNRAS.504..792C}). Unlike AT2020ohl, the other radio
 luminous TDEs became optically thin at these radio frequencies by
 300 days after the peak. This radio characteristic of AT2020ohl
 distinguishes it from the canonical radio luminous TDEs, and in fact, advocates
 its non-TDE origin. 
  
\subsection{The combined picture}\label{sec:dis_comb}
\begin{table*}
\centering
\caption{Parameters of JetSeT model}
\label{tab:SED_mod}
\begin{tabular}{lcccc} 
\hline
Parameter (unit) & Start value & Min value & Max value & Best-fit value (error) \\
\hline
$\gamma_{min}$&  10$^a$        &  $--$ &  $--$ & $--$ \\
$\gamma_{max}$& 5E+05$^a$      &  $--$ &  $--$ & $--$ \\
N (cm$^{-3}$) & 1000       &  50   & 5000  & 683.35 ($\pm330.76$) \\
$p$           & 2.5        & $-$10 & +10   & 1.52 ($\pm0.07$) \\
R$_{in}$/R$_{sh}$&   50$^a$    & $--$  & $--$  & $--$ \\
R$_{out}$/R$_{sh}$& 100$^a$    & $--$  & $--$  & $--$ \\
$\alpha_{\rm eff}$& 0.01       & 0.0   &  0.1  & 0.006 ($\pm0.005$) \\
M$_{BH}$ (M$_\odot$) & 1E+07 & 9E+06  & 1E+08 & 1.2E+07 ($\pm$3.9E+06) \\
L$_{disk}$ (erg\,s$^{-1}$)& 1E+43$^a$ &   $--$ &  $--$ & $--$ \\
R  (cm)   & 9E+14$^a$ &   $--$ &  $--$ & $--$ \\
R$_H$ (cm)& 1E+17$^a$ &   $--$ &  $--$ & $--$ \\
B  (Gauss)& 0.1 & 0.05 & 0.5 & 0.08 ($\pm$0.02) \\
$\theta$ ($\degree$) &  5$^a$ &  $--$ &  $--$ & $--$ \\
$\Gamma$  &  2$^a$  &   $--$ &  $--$ & $--$ \\
$z$       &  0.1671$^a$ &  $--$ &  $--$ & $--$ \\
\hline
\end{tabular}
\begin{tablenotes}
\item $^a$ Parameters, which are fixed, have only the `Start values' (2nd column). The
 parameters being varied have a start value (2nd column) with minimum and maximum limits 
 (3rd and 4th columns, respectively). The best-fit values of these parameters are given 
 in the 5th column.
 \end{tablenotes}
\end{table*}

 From multi-wavelength analysis, it was well  established that AT2020ohl was a
 slowly evolving transient.
 To understand
 the characteristic of its emission, the SED of the object has been modeled 
 using multi-wavelength broadband data. In this regard, We used the X-band and C-band
 radio data of the transient observed on +318 day, along with the X-ray data from 
 {\it Swift}/XRT, and host-subtracted NUV-optical data from {\it Swift}/UVOT 
 observed on the +322 day (observations are quasi-simultaneous). 
 The redshift corrected observed broadband SED is shown in Figure \ref{fig:SEDmod} (see the upper-right panel). We modelled the 
 broadband SED  using the `Jets SED modeler and fitting Tool' (JetSeT,
 \citealt{2006A&A...448..861M, 2009A&A...501..879T, 2011ApJ...739...66T,
 2020ascl.soft09001T}). It is important to note that the host of AT2020ohl is
 an S0 galaxy although had a mild pre-transient X-ray emission, showing mild AGN activity and no star-forming lines in its
 spectrum (\S\ref{sec:host}). The log of the ratio of its axes (i.e., major
 axis/minor axis) is {\it logr25}\footnote{Here the length of each axis is the
 distance from the center where surface brightness is reduced to 25
 mag/arcsec$^2$} = 0.16$\pm$0.05 \citep{2014A&A...570A..13M}, showing that
 it is a nearly face-on galaxy. Since the host is a mildly-active galaxy,
 we consider that there is neither a torus nor a narrow/broad-line regions
 surrounding the SMBH. This is also consistent with the spectra of the host.

 The observed SED is modelled assuming a very simple geometry of the
 system. During the transient activity, we assume a
 temporary accretion disk is produced (or there was already a 
 disk-like structure present in this system). This is the origin of the black body radiation, 
 which peaks in UV wavelength and maintaining a constant disk temperature.
 The overall geometry of the system is Blazar-like, without the
 presence of a torus and narrow \& broad line-forming regions. The non-thermal
 radiation is produced by a spherical homogeneous blob of leptonic plasma
 (e$^{\pm}$) entangled with magnetic fields, and it is moving along the jet,
 which is perpendicular to the plane of the disk. In the
 JetSeT environment, the system has a non-rotating SMBH of mass M$_{BH}$ at its center. 
 Therefore, as described above, the components of the emitted broad-band spectrum are 
 $-$ (1) a black body component produced by a Keplerian disk with an inner radius
 R$_{in}$ and outer radius R$_{out}$ respectively having a disk luminosity L$_{disk}$, and accretion efficiency $\alpha_{eff}$; and (2) a  
 non-thermal component, which is produced by the spherical blob
 having a radius R and at a distance R$_H$ from the SMBH. The emitter density
 within the blob is n, the magnetic field is B, and the bulk Lorentz factor of
 its material is $\Gamma$. The motion of the individual emission regions is 
 random and entangled with the magnetic field, causing a spread in the
 Lorentz-factor between a minimum ($\gamma_{min}$) and maximum
 ($\gamma_{max}$). The energy distribution of the particles in the blob is assumed to be a 
 power-law (i.e., n(E) $\propto$E$^{-p}$), and the jet points along a small angle $\theta$ with respect to the line of sight.

 Table \ref{tab:SED_mod} summarizes the initial and final fitted values of the 
 parameters. Since the redshift is well constrained its
 value is fixed. Given the limited number of constraints, 
 we have also frozen several other parameters in
 this fitting process. We explored the 
 number density of the emitter (n), the power-law index
 ($p$) of the synchrotron emitting plasma, the accretion efficiency ($\alpha_{eff}$)
 of the disk, the mass of the black hole (M$_{BH}$), and the flux density (B) of the
 magnetic field. The transient was more luminous in the NUV than the optical, indicating that the black body temperature is high and peaking at a lower wavelength. We fixed the L$_{disk}$ at $\sim10^{43}$ erg\,s$^{-1}$
 (which is consistent with the bolometric luminosity at comparable epoch) so
 that the black body emission from the disk peaks in the UV. We further assumed that
 a mildly relativistic outflow produces the non-thermal emission
 ($\Gamma\sim2$) and the $\theta\sim5\degree$. The values of all the other
 parameters (viz. $\gamma_{min}$, $\gamma_{max}$, R$_{in}$/R$_{sh}$ (where R$_{sh}$ is the ``Schwarzschild radius'' of the SMBH),
 R$_{out}$/R$_{sh}$, R, and R$_H$) are  assumed to be consistent with their
 typical values for a blazer system. The priors (initial-starting and fixed values)
 are tabulated in the first column of Table \ref{tab:SED_mod}, and 
 the best-fit parameters (posteriors) and their errors are tabulated in column 5.
 The model SED is shown in 
 Figure \ref{fig:SEDmod}, along with the posterior distribution of the parameters. 

 The non-thermal emission produced by the mild-relativistic outflow is synchrotron self-absorption (SSA) dominated in the radio frequencies and is optically thick. 
 However, it is optically thin at higher frequencies (e.g., X-ray). 
 We have also searched for the
 Gamma-ray counterpart from the $FmildERMI$ archive and nothing was detected,
 indicating the absence of inverse Comptonized photons.
 Thus the SED is consistent with a disk-like accretion along with a mild-relativistic outflow.
  
\subsection{Probable origin of AT2020ohl}\label{sec:dis_origin}
 We rule out the possibility of associating
stellar explosion or disruption phenomena with the transient event AT2020ohl/ASASSN-20hx 
from the multi-wavelength analysis.
 The absence of any regular supernova line in the spectra immediately dismisses
 a nuclear-SN origin of this event. The same argument is also primarily
 applicable to rejecting the possibility of a TDE. 
 
 The rate of the rise of AT2020ohl and its host's mass are not similar to those
 of canonical TDEs \citep{2022ApJ...930...12H}. On 
 the other hand, as discussed in
 \S\ref{sec:uvopt}, between +120 and +400 day, the temporal decay index of the
 UV-optical lightcurve was $-0.41\pm$0.02. A similar decay rate was also observed
 in the X-ray light curve, which is much flatter than the t$^{-5/3}$ law. These
 properties show that AT2020ohl is most likely not associated with a stellar
 disruption event under the gravity of SMBH.

 Nevertheless, the combined disc-wind models of TDEs \citep{2009MNRAS.400.2070S, 2011MNRAS.410..359L} can explain the
 UV-optical lightcurves to some extent. According to these models, the bound
 matters of the star form disk while the unbound debris form wind-like outflow,
 causing a relatively flatter post-maximum decay in NUV-optical luminosity, maintaining 
 a nature of t$^{-5/12}$. Although this decay index is consistent with the observations of AT2020ohl, there are large differences 
 between the observed and predicted X-ray luminosities and X-ray decay
 rates. While models predict a higher luminosity and roughly t$^{-5/3}$ decay
 law over a period of $\sim1000$ days after disruption, the observed X-ray luminosity of 
 AT2020ohl was much less than the NUV luminosity, and the
 corresponding X-ray decay rate was flatter than t$^{-5/3}$ law. Moreover, the
 origin of the X-ray is thermal by the \citet{2011MNRAS.410..359L} model,
 whereas, in \S\ref{sec:dis_comb}, we argued that it is completely non-thermal.

 From the UV-optical bolometric lightcurve, the migration of the radius-temperature relation toward a steady disk condition
 (\S\ref{sec:dis_uvop}), and the multi-band SED modelling (\S\ref{sec:dis_comb}), it
 is now established that an optically thick thermally-emitting accretion disk must be 
 present in this system. The variation in the UV-optical 
lightcurves and the non-standard behaviour of the radius-temperature relation at the early phases may be a manifestation of a dynamically evolving accretion disk. Since the base of the jet is supposed to be connected with the inner accretion disk, a variation in the X-ray light curve is also expected, if it originates due to the reprocessing of accretion disk photons. However, as discussed in \S\ref{sec:dis_comb},
 the X-ray and NUV-optical photons may have different origin and a strong correlation between light 
 curves in these wavebands is unexpected.
 
 A deeper investigation, like high-resolution multi-wavelength imaging, is
 required to probe the center of NGC6297 and to understand the origin of such
 ambiguous nuclear transients. It is noticeable that NGC6297 is an S0 galaxy
 which may be a post-merger remnant. The presence of an extended galaxy exponential disk profile 
 in this system (\S\ref{sec:host}) strengthens the post-merging scenario.
 The plausible existence of an LLAGN at the center of the host further
 strengthens the existence of a mild-accretion disk around the SMBH,
 which is otherwise not luminous. It is possible that in long periodical
 intervals,
 this disk either gets an additional matter supply from its host or the disk is 
 intervened by some other compact object (like a bare-SMBH of a merger), causing
 a sudden accretion and flaring activity of the disk in the UV-optical bands as 
 well as an outflow, which is the origin of the non-thermal X-ray and radio 
 emission. A similar possibility was studied in the BL-Lac
 object OJ287 based on its long-term optical lightcurve
 (e.g., \citealt{1988ApJ...325..628S}), radio VLBA observations (e.g.,
 \citealt{2012ApJ...747...63A, Gomez_2022}), and observations in other wavebands. Recently,
 high-spatial resolution ($\sim$0\farcs1) observations of UGC4211 $-$ a post-merger galaxy have revealed the existence of binary SMBH with
 separation of $\sim230$pc \citep{Koss_2023}. Noteworthy, VLBI
 observations have confirmed the existence of binary SMBHs, where the separation
 between two compact objects is as less as 7.3 pc (e.g.,
 \citealt{2017NatAs...1..727K} and references therein). Therefore, the existence
 of binary SMBH at the center of NGC6297 can not be ruled out, and certainly,
 high-spatial resolution observations are necessary to confirm this possibility. 
 
 In the JVLA B-array X-band radio image (Figure
 \ref{fig:OptXrayRadio}), the knot observed in the north-east direction from the
 center of the galaxy is at a projected angular separation of $1\farcs5$, which corresponds to a distance of $\sim265$pc
 ($\equiv$ 850 lightyears), which  
 implies that this knot is not causally related to AT2020ohl.
 Unfortunately, we could not obtain JVLA observations at other frequencies 
 (preferably in the C-band) with comparable spatial resolution so that the spectral
 index of the knot and, therefore, its age could be calculated. However, given that
 the X-band flux of the knot is comparable to that of the central emission
 (\S\ref{sec:radioevol}) if we assume that the knot was indeed associated with
 the central activity due to an event similar to AT2020ohl, it must happened
 more than 850 years ago.

 The transition between active and inactive galaxies and vice versa is not
 completely understood. Merging of galaxies may be a key process for transition
 from inactive to active state. Since S0-galaxies have gone through the merging
 process, the study of their central regions is important to probe the
 post-merger scenario. It will also shed light on the gas accumulation process
 at the centers of these systems. The discovery of AT2020ohl/ASASSN-20hx is an
 example in this regard. Although in this study we ruled out the association of
 stellar explosion or disruption phenomena, we could not precisely point out whether the
 sudden supply of matter to the pre-existing accretion disk triggered the event
 or it is due to a close interaction with another SMBH which was already a part
 of a binary SMBH system at the center of the host.  
 Multi-wavelength high-resolution imaging has been planned to probe this system
 in the near future.

\section*{Acknowledgements}
 We thank all the  observers of the 2 m  HCT (operated by the Indian Institute 
 of Astrophysics), who kindly provided part of their observing time for the
 optical spectroscopic observations of this transient. We acknowledge the use of
 public data from the Swift data archive. This research has made use of data
 obtained from the Chandra Data Archive and the Chandra Source Catalog, and
 software provided by the Chandra X-ray Center (CXC) in the application
 packages CIAO and Sherpa. The National Radio Astronomy Observatory is a
 facility of the National Science Foundation operated under cooperative
 agreement by Associated Universities, Inc. This work has made use of
 the NASA Astrophysics Data System and the NASA/IPAC Extragalactic Database
 (NED) which is operated by Jet Propulsion Laboratory, California Institute of 
 Technology, under contract with the National Aeronautics and Space
 Administration. RR acknowledges IUCAA for providing access to their
 High-Performance Computing (HPC) facility for the analysis of JVLA data. Manipal 
 Centre for Natural Sciences, Centre of Excellence, Manipal Academy of Higher Education 
 (MAHE) is acknowledged for facilities and support. SN acknowledges support by the 
 Science \& Engineering Research Board (SERB), a statutory body of Department of 
 Science \& Technology (DST), Government of India (SERB-SURE Grant, File Number: 
 SUR/2022/003864).
 This work has used data from the Sloan Digital Sky Survey (SDSS). Funding for
 the Sloan Digital Sky Survey V has been provided by the Alfred P. Sloan
 Foundation, the Heising-Simons Foundation, the National Science Foundation,
 and the Participating Institutions. SDSS acknowledges support and resources
 from the Center for High-Performance Computing at the University of Utah. The
 SDSS web site is \url{www.sdss.org}.
 SDSS is managed by the Astrophysical Research Consortium for the Participating
 Institutions of the SDSS Collaboration, including the Carnegie Institution for
 Science, Chilean National Time Allocation Committee (CNTAC) ratified
 researchers, the Gotham Participation Group, Harvard University, Heidelberg
 University, The Johns Hopkins University, L’Ecole polytechnique
 f\'{e}d\'{e}rale de Lausanne (EPFL), Leibniz-Institut f\''{u}r Astrophysik
 Potsdam (AIP), Max-Planck-Institut f\''{u}r Astronomie (MPIA Heidelberg),
 Max-Planck-Institut f\''{u}r Extraterrestrische Physik (MPE), Nanjing
 University, National Astronomical Observatories of China (NAOC), New Mexico
 State University, The Ohio State University, Pennsylvania State University,
 Smithsonian Astrophysical Observatory, Space Telescope Science Institute
 (STScI), the Stellar Astrophysics Participation Group, Universidad Nacional
 Aut\'{o}noma de M\'{e}xico, University of Arizona, University of Colorado
 Boulder, University of Illinois at Urbana-Champaign, University of Toronto,
 University of Utah, University of Virginia, Yale University, and Yunnan
 University. 

\section*{Data Availability}
 The data used in this article will be shared on request to the corresponding
 author. The VLA data underlying this article is available from the NRAO 
 Science Data Archive (\url{https://archive.nrao.edu/archive/advquery.jsp}).  
 The data obtained from {\it Swift} and CHANDRA underlying this article are
 available at HEASARC
 archive (\url{https://heasarc.gsfc.nasa.gov/docs/archive.html}). The data from 
 ATLAS is available at ATLAS forced photometry archive
 (\url{https://fallingstar-data.com/forcedphot/}). The spectroscopic data 
 obtained from HCT for this project will be shared on reasonable request to the
 corresponding author.


\bibliographystyle{mnras}
\bibliography{references} 




\appendix

\section{Some extra material}
\begin{table*}
\centering
\caption{Journal of spectroscopic observations of AT2020ohl}
\label{tab:speclog}
\begin{tabular}{lc cc cc c cc}
\hline
UT Date & JD & Phase$^a$ & Range & Telescope & Grating & Exposure & S/N$^b$& Slit width\\

    (yy-mm-dd)& 2450000+& (days)& \mum&     & (gr mm$^{-1}$)&(s)&(pix$^{-1}$)&  (\arcsec) \\
  \hline

   2020-08-31&9093.14&+68&0.36$-$0.9&HCT/HFOSC2&GR7+GR8&2700/2700&45&1.5\\
   2020-09-08&9100.11&+76&0.36$-$0.9&HCT/HFOSC2&GR7+GR8&2700/2700&43&1.5\\
   2020-09-29&9122.11&+97&0.36$-$0.9&HCT/HFOSC2&GR7+GR8&2700/2700&34&1.5\\
   2020-10-16&9138.70&+113&0.36$-$0.9&HCT/HFOSC2&GR7+GR8&2700/2700&34&1.5\\
   2020-12-02&9186.05&+160&0.36$-$0.9&HCT/HFOSC2&GR7+GR8&2700/2700&35&1.5\\
   2021-03-27&9301.30&+273&0.36$-$0.8&HCT/HFOSC2&GR7    &2700     &38&1.5\\
   2021-10-19&9507.14&+475&0.36$-$0.8&HCT/HFOSC2&GR7    &1800     &35&1.5\\
  \hline
   2000-05-30$^c$&1695 &-7328&0.38$-$0.9&SDSS/SPEC2& Blue/Red & 3600&48&3 (fiber/ID 485)\\
  \hline
\end{tabular}

\begin{tablenotes}
\item $^a$ With reference to the explosion epoch JD 2459023.3
\item $^b$ At 0.6 \mum.\\
\item $^c$ This epoch marks the pre-transient spectroscopic observation of the host by
 the SDSS survey. 
\end{tablenotes} 

\end{table*}

  \begin{table*}
  \caption{Log of radio observation of AT2020ohl$^a$ from JVLA in C (4-8 GHz) \&
  X (8-12 GHz).}
  \centering
  \label{tab:vladata}
  \begin{tabular}{ccccccc}
  \hline
  UT Date   &JD      &Phase$^b$&Frequency&Array&Flux     &Flux error$^c$\\
  (yy/mm/dd)&2450000+& (day)   & (GHz)   &     &($\mu$Jy)& ($\mu$Jy)\\
  \hline
  2020/10/29&9152.46& +128&10 (X-band)&B&33.5    &3.7\\
  2021/05/19&9353.64& +330&10 (X-band)&D&47      &14.7\\
  2021/05/24&9358.63& +335&05 (C-band)&D&$\sim$30& -- \\
  \hline
  \end{tabular}
\begin{tablenotes}
\item $^{a}${Here, only the fluxes of the central emission from NGC6297 have been
 tabulated}
\item $^{b}${With reference to the explosion epoch JD 2459023.3}
\item $^{c}${The upper limit has no error bars}
\end{tablenotes}
  \end{table*}

\begin{table*}
\centering
\caption{$Swift$/UVOT photometry of AT2020ohl$^a$}
\label{tab:phot_uvot}
\begin{tabular}{ccccccccc}
\hline
UT Date      & JD        & Phase$^b$  & $uw2$ & $um2$ & $uw1$ & $uuu$ & $ubb$ & $uvv$ \\
(yyyy/mm/dd) & (2450000+)& (days) & (mag) & (mag) & (mag) & (mag) & (mag) & (mag) \\
\hline
2020/07/19&9049.68&+26 &15.94$\pm$0.06&16.00$\pm$0.05&15.93$\pm$0.05&15.62$\pm$0.04&15.12$\pm$0.04&14.61$\pm$0.04\\
2020/07/20&9051.14&+27 &15.92$\pm$0.06&15.91$\pm$0.05&15.85$\pm$0.05&15.64$\pm$0.04&15.15$\pm$0.04&14.59$\pm$0.04\\
2020/07/21&9052.02&+28 &15.92$\pm$0.06&15.92$\pm$0.05&15.88$\pm$0.05&15.66$\pm$0.05&15.12$\pm$0.04&14.64$\pm$0.04\\
2020/07/22&9053.23&+29 &15.90$\pm$0.06&15.95$\pm$0.06&15.88$\pm$0.05&15.64$\pm$0.05&15.12$\pm$0.05&14.58$\pm$0.05\\
2020/07/23&9053.66&+30 &15.98$\pm$0.06&15.95$\pm$0.05&15.91$\pm$0.05&15.62$\pm$0.05&15.13$\pm$0.04&14.58$\pm$0.04\\
2020/07/25&9055.59&+32 &15.99$\pm$0.06&15.96$\pm$0.05&15.93$\pm$0.05&15.69$\pm$0.05&15.15$\pm$0.05&14.55$\pm$0.05\\
2020/07/27&9057.72&+34 &16.01$\pm$0.06&15.97$\pm$0.05&15.92$\pm$0.05&15.69$\pm$0.05&15.17$\pm$0.04&14.62$\pm$0.04\\
2020/07/29&9060.43&+37 &15.95$\pm$0.06&15.89$\pm$0.05&15.87$\pm$0.05&15.66$\pm$0.05&15.15$\pm$0.04&14.61$\pm$0.04\\
2020/07/31&9061.90&+38 &16.03$\pm$0.06&15.99$\pm$0.05&15.87$\pm$0.05&15.61$\pm$0.05&15.13$\pm$0.04&14.64$\pm$0.04\\
2020/08/02&9063.76&+40 &16.14$\pm$0.06&16.04$\pm$0.05&15.93$\pm$0.05&15.60$\pm$0.05&15.15$\pm$0.04&14.62$\pm$0.04\\
2020/08/04&9065.94&+42 &16.10$\pm$0.06&16.02$\pm$0.05&15.94$\pm$0.05&15.59$\pm$0.05&15.19$\pm$0.04&14.61$\pm$0.04\\
2020/08/06&9067.87&+44 &15.97$\pm$0.06&15.88$\pm$0.05&15.84$\pm$0.05&15.56$\pm$0.05&15.13$\pm$0.04&14.62$\pm$0.04\\
2020/08/08&9070.03&+46 &15.98$\pm$0.06&15.90$\pm$0.05&15.87$\pm$0.05&15.55$\pm$0.05&15.11$\pm$0.04&14.59$\pm$0.04\\
2020/08/10&9071.75&+48 &15.96$\pm$0.06&15.92$\pm$0.05&15.82$\pm$0.05&15.57$\pm$0.05&15.10$\pm$0.04&14.61$\pm$0.04\\
2020/08/16&9078.14&+54 &15.99$\pm$0.06&15.92$\pm$0.05&15.88$\pm$0.05&15.57$\pm$0.05&15.10$\pm$0.04&14.69$\pm$0.04\\
2020/08/18&9080.13&+56 &16.01$\pm$0.06&15.98$\pm$0.05&15.89$\pm$0.05&15.61$\pm$0.05&15.13$\pm$0.05&14.62$\pm$0.05\\
2020/08/20&9082.29&+58 &16.13$\pm$0.06&16.03$\pm$0.05&15.94$\pm$0.05&15.64$\pm$0.05&15.08$\pm$0.04&14.58$\pm$0.04\\
2020/08/22&9084.29&+60 &16.03$\pm$0.06&15.97$\pm$0.05&15.89$\pm$0.05&15.61$\pm$0.05&15.16$\pm$0.05&14.57$\pm$0.04\\
2020/08/24&9085.98&+62 &16.09$\pm$0.06&16.04$\pm$0.05&15.95$\pm$0.05&15.62$\pm$0.05&15.09$\pm$0.04&14.63$\pm$0.04\\
2020/08/26&9087.97&+64 &16.10$\pm$0.06&16.01$\pm$0.05&15.90$\pm$0.05&15.61$\pm$0.05&15.13$\pm$0.04&14.59$\pm$0.04\\
2020/08/28&9090.09&+66 &16.12$\pm$0.06&16.05$\pm$0.05&15.89$\pm$0.05&15.57$\pm$0.05&15.11$\pm$0.04&14.61$\pm$0.04\\
2020/08/30&9092.02&+68 &16.08$\pm$0.06&16.03$\pm$0.05&15.93$\pm$0.05&15.65$\pm$0.05&15.10$\pm$0.04&14.63$\pm$0.04\\
2020/09/01&9093.68&+70 &16.21$\pm$0.06&     $--$     &15.95$\pm$0.05&15.67$\pm$0.05&15.24$\pm$0.05&14.67$\pm$0.05\\
2020/09/03&9095.96&+72 &16.06$\pm$0.06&16.06$\pm$0.05&15.94$\pm$0.05&15.57$\pm$0.05&15.15$\pm$0.04&14.57$\pm$0.04\\
2020/09/05&9098.06&+74 &16.13$\pm$0.06&16.11$\pm$0.05&15.97$\pm$0.05&15.64$\pm$0.05&15.12$\pm$0.04&14.59$\pm$0.04\\
2020/09/07&9099.94&+76 &16.17$\pm$0.06&16.08$\pm$0.05&16.00$\pm$0.05&15.68$\pm$0.05&15.17$\pm$0.04&14.66$\pm$0.04\\
2020/09/09&9101.80&+78 &16.07$\pm$0.06&16.03$\pm$0.05&16.01$\pm$0.05&15.67$\pm$0.05&15.18$\pm$0.04&14.65$\pm$0.04\\
2020/09/11&9103.73&+80 &16.19$\pm$0.06&16.10$\pm$0.05&15.97$\pm$0.05&15.65$\pm$0.05&15.16$\pm$0.04&14.65$\pm$0.04\\
2020/09/13&9106.11&+82 &16.11$\pm$0.06&16.05$\pm$0.05&15.93$\pm$0.05&15.65$\pm$0.05&15.17$\pm$0.04&14.65$\pm$0.04\\
2020/09/15&9107.94&+84 &16.09$\pm$0.06&16.03$\pm$0.05&15.92$\pm$0.05&15.68$\pm$0.05&15.13$\pm$0.05&14.58$\pm$0.04\\
2020/09/17&9109.70&+86 &16.11$\pm$0.06&16.06$\pm$0.05&15.97$\pm$0.05&15.59$\pm$0.05&15.15$\pm$0.04&14.65$\pm$0.05\\
2020/09/19&9111.56&+88 &16.11$\pm$0.06&16.02$\pm$0.05&15.98$\pm$0.05&15.66$\pm$0.05&15.10$\pm$0.04&14.64$\pm$0.04\\
2020/09/20&9113.15&+89 &16.11$\pm$0.06&16.07$\pm$0.05&16.00$\pm$0.05&15.65$\pm$0.05&15.08$\pm$0.04&14.61$\pm$0.04\\
2020/09/23&9115.77&+92 &16.12$\pm$0.06&16.06$\pm$0.05&15.94$\pm$0.05&15.64$\pm$0.04&15.16$\pm$0.04&14.65$\pm$0.04\\
2020/09/25&9118.06&+94 &16.21$\pm$0.06&16.16$\pm$0.05&15.99$\pm$0.05&15.65$\pm$0.05&15.15$\pm$0.04&14.73$\pm$0.04\\
2020/09/27&9119.56&+96 &16.05$\pm$0.06&15.96$\pm$0.06&15.92$\pm$0.06&15.61$\pm$0.05&15.15$\pm$0.05&14.62$\pm$0.05\\
2020/10/08&9131.49&+108&16.04$\pm$0.06&15.99$\pm$0.05&15.86$\pm$0.05&15.56$\pm$0.05&15.08$\pm$0.04&14.58$\pm$0.04\\
2020/10/14&9136.96&+113&16.03$\pm$0.06&15.98$\pm$0.05&15.85$\pm$0.05&15.51$\pm$0.05&15.09$\pm$0.05&14.59$\pm$0.05\\
2020/10/19&9141.91&+118&16.15$\pm$0.06&16.06$\pm$0.05&15.97$\pm$0.05&15.67$\pm$0.05&15.13$\pm$0.04&14.61$\pm$0.04\\
2020/10/20&9143.34&+120&16.12$\pm$0.06&16.04$\pm$0.05&15.92$\pm$0.05&15.60$\pm$0.05&15.17$\pm$0.05&14.59$\pm$0.04\\
2020/10/26&9149.08&+125&16.17$\pm$0.06&16.08$\pm$0.05&16.04$\pm$0.05&15.69$\pm$0.04&15.15$\pm$0.04&14.55$\pm$0.06\\
2020/10/29&9152.16&+128&16.24$\pm$0.06&16.10$\pm$0.05&16.06$\pm$0.05&15.65$\pm$0.04&15.13$\pm$0.04&14.58$\pm$0.04\\
2020/11/02&9156.11&+132&16.25$\pm$0.06&16.18$\pm$0.05&16.08$\pm$0.05&15.71$\pm$0.05&15.17$\pm$0.04&14.63$\pm$0.04\\
2020/11/07&9161.11&+137&16.13$\pm$0.06&16.06$\pm$0.05&15.99$\pm$0.05&15.64$\pm$0.05&15.15$\pm$0.04&14.60$\pm$0.04\\
2020/11/12&9166.32&+143&16.19$\pm$0.06&16.10$\pm$0.05&16.03$\pm$0.05&15.65$\pm$0.05&15.17$\pm$0.04&14.64$\pm$0.04\\
2020/11/17&9170.71&+147&16.26$\pm$0.06&16.16$\pm$0.05&16.08$\pm$0.05&15.72$\pm$0.05&15.19$\pm$0.05&14.61$\pm$0.05\\
2020/11/28&9182.21&+158&16.28$\pm$0.06&16.21$\pm$0.05&16.08$\pm$0.05&15.72$\pm$0.05&15.24$\pm$0.05&14.62$\pm$0.05\\
2020/12/02&9186.16&+162&16.34$\pm$0.06&16.20$\pm$0.05&16.13$\pm$0.05&15.78$\pm$0.05&15.20$\pm$0.04&14.65$\pm$0.05\\
2020/12/07&9190.78&+167&16.27$\pm$0.06&16.22$\pm$0.05&16.12$\pm$0.05&15.79$\pm$0.05&15.22$\pm$0.04&14.66$\pm$0.04\\
2020/12/12&9195.72&+172&16.39$\pm$0.06&16.30$\pm$0.05&16.18$\pm$0.05&15.89$\pm$0.05&15.26$\pm$0.04&14.72$\pm$0.04\\
2020/12/22&9206.18&+182&16.35$\pm$0.06&16.26$\pm$0.06&16.16$\pm$0.05&15.77$\pm$0.05&15.24$\pm$0.04&14.67$\pm$0.06\\
2020/12/27&9210.70&+187&16.32$\pm$0.06&16.22$\pm$0.05&16.13$\pm$0.05&15.84$\pm$0.05&15.25$\pm$0.04&14.72$\pm$0.04\\
2021/02/01&9246.99&+223&16.47$\pm$0.06&16.37$\pm$0.05&16.22$\pm$0.05&15.90$\pm$0.05&15.27$\pm$0.05&14.63$\pm$0.05\\
2021/02/15&9260.84&+237&16.54$\pm$0.06&16.46$\pm$0.05&16.31$\pm$0.06&15.91$\pm$0.06&15.28$\pm$0.05&14.71$\pm$0.05\\
2021/02/19&9265.01&+241&16.54$\pm$0.07&16.35$\pm$0.06&16.29$\pm$0.06&15.91$\pm$0.06&15.30$\pm$0.06&14.70$\pm$0.06\\
2021/03/02&9275.56&+252&16.47$\pm$0.06&16.38$\pm$0.06&16.31$\pm$0.06&15.92$\pm$0.06&15.28$\pm$0.06&14.70$\pm$0.06\\
2021/03/15&9288.59&+265&16.52$\pm$0.07&16.44$\pm$0.06&16.53$\pm$0.07&15.93$\pm$0.07&15.35$\pm$0.07&14.86$\pm$0.08\\
2021/03/18&9291.70&+268&16.61$\pm$0.06&16.55$\pm$0.06&16.36$\pm$0.06&15.95$\pm$0.06&15.33$\pm$0.05&14.73$\pm$0.05\\
2021/03/19&9292.70&+269&16.53$\pm$0.07&16.48$\pm$0.06&16.27$\pm$0.06&15.88$\pm$0.06&15.24$\pm$0.06&14.64$\pm$0.06\\
2021/03/29&9302.79&+279&16.60$\pm$0.06&16.58$\pm$0.05&16.45$\pm$0.05&15.92$\pm$0.05&15.28$\pm$0.05&14.67$\pm$0.05\\
2021/04/13&9317.59&+294&16.71$\pm$0.07&16.58$\pm$0.06&16.47$\pm$0.06&16.10$\pm$0.06&15.34$\pm$0.05&14.73$\pm$0.06\\
\hline
\end{tabular}
\begin{tablenotes}
 \item $^{a}$ The magnitude of the transient at different UVOT bands have been determined after subtracting the host flux at the transient location (see \S\ref{sec:host}).
 \item $^b$ With reference to the explosion epoch JD 2459023.3
 \end{tablenotes}
\end{table*}

\begin{table*}
\centering
\contcaption{Continuation of Table \ref{tab:phot_uvot}}
\begin{tabular}{ccccccccc}
\hline
UT Date      & JD        & Phase  & $uw2$ & $um2$ & $uw1$ & $uuu$ & $ubb$ & $uvv$ \\
(yyyy/mm/dd) & (2450000+)& (days) & (mag) & (mag) & (mag) & (mag) & (mag) & (mag) \\
\hline
2021/04/15&9320.43&+297&16.58$\pm$0.07&16.47$\pm$0.06&16.36$\pm$0.07&16.13$\pm$0.07&15.28$\pm$0.06&14.86$\pm$0.07\\
2021/05/12&9346.83&+323&16.76$\pm$0.06&16.58$\pm$0.05&16.50$\pm$0.05&16.07$\pm$0.05&15.29$\pm$0.05&14.80$\pm$0.05\\
2021/06/03&9369.03&+345&16.74$\pm$0.06&16.66$\pm$0.06&16.57$\pm$0.06&16.07$\pm$0.06&15.35$\pm$0.05&14.82$\pm$0.05\\
2021/06/24&9389.56&+366&16.81$\pm$0.07&16.47$\pm$0.14&16.54$\pm$0.06&16.16$\pm$0.06&15.31$\pm$0.05&14.72$\pm$0.06\\
2021/07/21&9417.01&+393&16.85$\pm$0.06&16.77$\pm$0.05&16.56$\pm$0.05&16.14$\pm$0.05&15.32$\pm$0.05&14.75$\pm$0.05\\
2021/08/14&9441.12&+417&16.70$\pm$0.06&16.59$\pm$0.05&16.48$\pm$0.05&16.07$\pm$0.05&15.29$\pm$0.04&14.73$\pm$0.04\\
2021/08/18&9445.33&+422&16.99$\pm$0.06&16.90$\pm$0.05&16.71$\pm$0.05&16.21$\pm$0.05&15.35$\pm$0.04&14.74$\pm$0.04\\
2021/09/15&9473.10&+449&16.60$\pm$0.06&16.63$\pm$0.05&16.49$\pm$0.05&16.07$\pm$0.05&15.34$\pm$0.04&14.77$\pm$0.04\\
2021/10/06&9494.32&+471&16.65$\pm$0.06&16.58$\pm$0.05&16.51$\pm$0.05&16.09$\pm$0.05&15.32$\pm$0.04&14.74$\pm$0.04\\
2021/10/13&9501.18&+477&16.63$\pm$0.06&16.60$\pm$0.05&16.52$\pm$0.05&16.05$\pm$0.05&15.35$\pm$0.04&14.77$\pm$0.04\\
2021/10/27&9515.25&+491&16.61$\pm$0.06&16.56$\pm$0.05&16.42$\pm$0.05&16.08$\pm$0.05&15.35$\pm$0.04&14.80$\pm$0.04\\
2021/11/10&9529.21&+505&16.56$\pm$0.06&16.48$\pm$0.05&16.46$\pm$0.05&16.15$\pm$0.05&15.40$\pm$0.04&14.77$\pm$0.04\\
2021/11/24&9543.43&+520&16.76$\pm$0.06&16.69$\pm$0.05&16.58$\pm$0.05&16.18$\pm$0.05&15.37$\pm$0.04&14.73$\pm$0.04\\
2021/12/10&9559.38&+536&16.61$\pm$0.06&16.48$\pm$0.06&16.42$\pm$0.05&16.10$\pm$0.05&15.33$\pm$0.04&14.80$\pm$0.04\\
2021/12/17&9566.25&+542&16.50$\pm$0.06&16.44$\pm$0.05&16.38$\pm$0.05&16.07$\pm$0.05&15.37$\pm$0.04&14.74$\pm$0.04\\
2021/12/24&9573.14&+549&16.53$\pm$0.06&16.46$\pm$0.05&16.42$\pm$0.05&16.11$\pm$0.05&15.35$\pm$0.04&14.77$\pm$0.04\\
2021/12/31&9580.31&+557&16.56$\pm$0.06&16.42$\pm$0.05&16.37$\pm$0.05&16.09$\pm$0.05&15.36$\pm$0.04&14.77$\pm$0.04\\
2022/01/14&9594.09&+570&16.53$\pm$0.06&16.52$\pm$0.05&16.44$\pm$0.05&16.14$\pm$0.05&15.44$\pm$0.05&14.77$\pm$0.05\\
2022/04/17&9686.70&+663&16.87$\pm$0.06&16.74$\pm$0.06&16.69$\pm$0.06&16.23$\pm$0.05&     $--$     &     $--$     \\
2022/05/01&9701.48&+678&16.88$\pm$0.06&16.77$\pm$0.06&16.72$\pm$0.06&16.22$\pm$0.05&     $--$     &     $--$     \\
2022/05/05&9704.76&+681&16.96$\pm$0.06&16.86$\pm$0.06&16.77$\pm$0.06&16.26$\pm$0.05&     $--$     &     $--$     \\
2022/05/15&9714.80&+691&17.00$\pm$0.06&16.93$\pm$0.06&16.79$\pm$0.05&16.31$\pm$0.05&     $--$     &     $--$     \\
2022/05/29&9728.62&+705&17.05$\pm$0.07&16.89$\pm$0.07&16.75$\pm$0.07&16.32$\pm$0.06&     $--$     &     $--$     \\
2022/06/02&9732.86&+709&16.91$\pm$0.06&16.83$\pm$0.06&16.74$\pm$0.06&16.36$\pm$0.05&     $--$     &     $--$     \\
2022/06/15&9746.47&+723&17.22$\pm$0.07&17.08$\pm$0.07&16.93$\pm$0.06&16.28$\pm$0.05&     $--$     &     $--$     \\
2022/06/16&9746.94&+723&16.92$\pm$0.06&16.85$\pm$0.06&16.73$\pm$0.06&16.29$\pm$0.05&     $--$     &     $--$     \\
2022/06/26&9757.19&+733&16.84$\pm$0.06&16.79$\pm$0.06&16.67$\pm$0.05&16.29$\pm$0.05&     $--$     &     $--$     \\
2022/07/10&9771.00&+747&16.93$\pm$0.06&16.86$\pm$0.06&16.71$\pm$0.05&16.23$\pm$0.05&     $--$     &     $--$     \\
\hline
\end{tabular}
\end{table*}

\begin{table*}
\centering
\caption{$Swift$/XRT observations of AT2020ohl}
\label{tab:phot_xrt}
\begin{tabular}{ccccc}
\hline
UT Date      & JD        & Flux (0.3-10 KeV)                  & Spectral Index$^a$ & Hardness Ratio$^b$ \\
(yyyy/mm/dd) & (2450000+)& 10$^{-12}$ erg\,cm$^{-2}$\,s$^{-1}$&                    &                    \\
\hline
 2020/07/18&9049.178&3.23$\pm$0.26&2.34$\pm$0.12&0.170$\pm$0.037\\
 2020/07/20&9050.637&3.36$\pm$0.39&2.01$\pm$0.07&0.244$\pm$0.048\\
 2020/07/20&9051.486&2.19$\pm$0.30&2.09$\pm$0.11&0.203$\pm$0.058\\
 2020/07/22&9052.729&2.20$\pm$0.30&2.36$\pm$0.19&0.147$\pm$0.057\\
 2020/07/22&9053.161&2.58$\pm$0.26&2.66$\pm$0.15&0.104$\pm$0.030\\
 2020/07/24&9055.085&3.0 $\pm$0.40&2.19$\pm$0.09&0.251$\pm$0.067\\
 2020/07/26&9057.217&2.03$\pm$0.29&2.11$\pm$0.19&0.236$\pm$0.078\\
 2020/07/29&9059.930&2.35$\pm$0.20&2.40$\pm$0.12&0.154$\pm$0.036\\
 2020/07/30&9061.398&2.34$\pm$0.24&2.29$\pm$0.21&0.210$\pm$0.044\\
 2020/08/01&9063.257&2.11$\pm$0.18&2.36$\pm$0.13&0.225$\pm$0.050\\
 2020/08/03&9065.447&2.30$\pm$0.23&2.22$\pm$0.13&0.170$\pm$0.043\\
 2020/08/05&9067.369&2.08$\pm$0.17&2.18$\pm$0.12&0.213$\pm$0.046\\
 2020/08/08&9069.530&2.09$\pm$0.25&2.10$\pm$0.18&0.196$\pm$0.047\\
 2020/08/09&9071.254&2.29$\pm$0.22&2.18$\pm$0.13&0.240$\pm$0.056\\
 2020/08/16&9077.643&1.89$\pm$0.19&2.09$\pm$0.14&0.256$\pm$0.065\\
 2020/08/18&9079.635&1.64$\pm$0.20&2.35$\pm$0.17&0.153$\pm$0.054\\
 2020/08/20&9081.791&1.65$\pm$0.21&2.37$\pm$0.26&0.195$\pm$0.051\\
 2020/08/22&9083.790&1.90$\pm$0.23&2.53$\pm$0.19&0.162$\pm$0.054\\
 2020/08/23&9085.484&2.20$\pm$0.32&2.17$\pm$0.10&0.187$\pm$0.049\\
 2020/08/25&9087.470&1.44$\pm$0.15&2.50$\pm$0.16&0.174$\pm$0.050\\
 2020/08/28&9089.595&2.30$\pm$0.21&2.21$\pm$0.13&0.198$\pm$0.045\\
 2020/08/30&9091.518&2.27$\pm$0.19&2.10$\pm$0.12&0.224$\pm$0.047\\
 2020/08/31&9093.277&3.2 $\pm$0.40&1.83$\pm$0.09&0.278$\pm$0.057\\
 2020/09/02&9095.457&2.03$\pm$0.22&2.12$\pm$0.13&0.190$\pm$0.049\\
 2020/09/05&9097.555&2.63$\pm$0.29&2.09$\pm$0.09&0.249$\pm$0.049\\
 2020/09/06&9099.440&2.13$\pm$0.35&1.90$\pm$0.09&0.331$\pm$0.080\\
 2020/09/08&9101.299&2.49$\pm$0.19&2.20$\pm$0.11&0.204$\pm$0.041\\
 2020/09/10&9103.228&2.91$\pm$0.20&2.31$\pm$0.10&0.172$\pm$0.033\\
 2020/09/13&9105.609&2.28$\pm$0.20&2.12$\pm$0.11&0.296$\pm$0.061\\
 2020/09/14&9107.443&2.82$\pm$0.25&2.25$\pm$0.13&0.166$\pm$0.042\\
 2020/09/16&9109.202&1.66$\pm$0.21&2.19$\pm$0.18&0.272$\pm$0.090\\
 2020/09/18&9111.062&2.48$\pm$0.23&2.16$\pm$0.12&0.206$\pm$0.044\\
 2020/09/20&9112.653&2.77$\pm$0.21&2.16$\pm$0.11&0.242$\pm$0.045\\
 2020/09/22&9115.272&2.10$\pm$0.19&2.11$\pm$0.12&0.196$\pm$0.040\\
 2020/09/25&9117.563&2.68$\pm$0.19&2.40$\pm$0.10&0.151$\pm$0.029\\
 2020/09/26&9119.093&2.59$\pm$0.24&2.22$\pm$0.13&0.193$\pm$0.046\\
 2020/10/08&9130.991&1.53$\pm$0.15&2.45$\pm$0.16&0.147$\pm$0.043\\
 2020/10/13&9136.463&2.0 $\pm$0.30&2.2 $\pm$0.40&0.166$\pm$0.059\\
 2020/10/18&9141.403&1.38$\pm$0.17&2.74$\pm$0.21&0.108$\pm$0.049\\
 2020/10/20&9142.838&2.36$\pm$0.25&2.31$\pm$0.14&0.220$\pm$0.055\\
 2020/10/26&9148.575&2.20$\pm$0.23&2.29$\pm$0.14&0.109$\pm$0.034\\
 2020/10/29&9151.660&1.54$\pm$0.16&2.29$\pm$0.15&0.175$\pm$0.056\\
 2020/11/02&9155.616&2.15$\pm$0.24&2.36$\pm$0.19&0.154$\pm$0.038\\
 2020/11/07&9160.613&1.69$\pm$0.15&2.36$\pm$0.13&0.223$\pm$0.055\\
 2020/11/12&9165.822&2.10$\pm$0.20&2.12$\pm$0.13&0.229$\pm$0.051\\
 2020/11/16&9170.205&2.5 $\pm$0.40&1.85$\pm$0.08&0.268$\pm$0.065\\
 2020/11/28&9181.680&0.89$\pm$0.14&2.31$\pm$0.21&0.509$\pm$0.202\\
 2020/12/02&9185.658&1.55$\pm$0.20&2.42$\pm$0.18&0.109$\pm$0.042\\
 2020/12/06&9190.274&1.50$\pm$0.21&2.32$\pm$0.12&0.148$\pm$0.047\\
 2020/12/11&9195.217&1.42$\pm$0.16&2.25$\pm$0.16&0.292$\pm$0.075\\
 2020/12/22&9205.675&1.31$\pm$0.22&2.8 $\pm$0.50&0.076$\pm$0.043\\
 2020/12/26&9210.074&0.97$\pm$0.16&2.5 $\pm$0.40&0.197$\pm$0.080\\
 2021/01/31&9246.490&1.34$\pm$0.23&1.65$\pm$0.17&0.359$\pm$0.129\\
 2021/02/14&9260.073&1.31$\pm$0.35&1.99$\pm$0.24&0.427$\pm$0.210\\
 2021/03/01&9275.065&1.70$\pm$0.40&1.74$\pm$0.26&0.407$\pm$0.163\\
 2021/03/14&9288.088&1.60$\pm$0.40&2.34$\pm$0.27&0.184$\pm$0.104\\
 2021/03/17&9291.205&0.92$\pm$0.19&3.10$\pm$0.70&0.106$\pm$0.067\\
 2021/03/18&9292.199&0.80$\pm$0.19&2.50$\pm$0.30&0.272$\pm$0.182\\
 2021/03/28&9302.290&1.28$\pm$0.24&2.02$\pm$0.15&0.295$\pm$0.099\\
 2021/04/12&9317.089&1.15$\pm$0.20&1.99$\pm$0.18&0.295$\pm$0.140\\
\hline
\end{tabular}
\begin{tablenotes}
\item $^a$ The spectral index ($\alpha$) is defined by the formula F$_{Xray}\propto\nu^{\alpha}$.
\item $^b$ The hardness ratio (HR) is defined as the ratio of 2-10 keV and 0.3-2 keV flux.
 \end{tablenotes}
\end{table*}
\begin{table*}
\centering
\contcaption{Continuation of Table \ref{tab:phot_xrt}}
\begin{tabular}{ccccc}
\hline
UT Date      & JD        & Flux (0.3-10 KeV)                  & Spectral Index$^a$ & Hardness Ratio$^b$ \\
(yyyy/mm/dd) & (2450000+)& 10$^{-12}$ erg\,cm$^{-2}$\,s$^{-1}$&                    &                    \\
\hline
 2021/05/11&9346.298&1.59$\pm$0.22&1.96$\pm$0.17&0.357$\pm$0.099\\
 2021/06/03&9368.526&0.94$\pm$0.23&1.77$\pm$0.18&0.350$\pm$0.170\\
 2021/06/23&9389.054&0.80$\pm$0.17&2.00$\pm$0.21&0.163$\pm$0.130\\
 2021/07/20&9416.049&1.15$\pm$0.21&1.91$\pm$0.19&0.278$\pm$0.102\\
 2021/08/13&9440.422&0.77$\pm$0.17&2.06$\pm$0.20&0.309$\pm$0.122\\
 2021/08/18&9444.803&0.91$\pm$0.14&1.88$\pm$0.19&0.365$\pm$0.131\\
 2021/09/14&9472.473&0.70$\pm$0.19&1.92$\pm$0.22&0.312$\pm$0.131\\
 2021/10/06&9493.723&0.61$\pm$0.09&2.45$\pm$0.24&0.177$\pm$0.084\\
 2021/10/12&9500.419&0.77$\pm$0.15&2.70$\pm$0.50&0.080$\pm$0.053\\
 2021/10/27&9514.608&0.55$\pm$0.12&1.97$\pm$0.15&0.340$\pm$0.142\\
 2021/11/09&9528.471&0.73$\pm$0.14&2.42$\pm$0.28&0.095$\pm$0.053\\
 2021/11/24&9542.890&0.83$\pm$0.12&2.29$\pm$0.22&0.235$\pm$0.086\\
 2021/12/10&9558.810&0.96$\pm$0.20&1.96$\pm$0.22&0.222$\pm$0.090\\
 2021/12/17&9565.588&0.84$\pm$0.07&2.20$\pm$0.13&0.291$\pm$0.107\\
 2021/12/23&9572.402&0.88$\pm$0.12&1.98$\pm$0.17&0.145$\pm$0.083\\
 2021/12/31&9579.766&0.31$\pm$0.09&1.90$\pm$0.40&0.558$\pm$0.374\\
 2022/05/14&9714.201&0.60$\pm$0.15&2.28$\pm$0.16&0.063$\pm$0.080\\
 2022/06/01&9732.350&0.86$\pm$0.35&2.10$\pm$0.48&0.131$\pm$0.106\\
 2022/07/09&9770.021&0.41$\pm$0.12&2.06$\pm$0.21&0.299$\pm$0.182\\
 2022/11/03&9886.558&0.58$\pm$0.14&2.60$\pm$0.50&0.223$\pm$0.129\\
\hline
\end{tabular}
\end{table*}


\bsp	
\label{lastpage}
\end{document}